\documentclass[showpacs,twocolumn,pra,superscriptaddress,notitlepage]{revtex4-1}
\usepackage{qcircuit}
\usepackage[dvips]{graphicx}
\usepackage{amsmath,amssymb,amsthm,mathrsfs,amsfonts,dsfont}
\usepackage{subfigure, epsfig}
\usepackage{braket}
\usepackage{bm}
\usepackage{enumerate}
\usepackage{color}
\usepackage{comment}
\usepackage[colorlinks = true]{hyperref}

\usepackage{newtxtext,newtxmath}

\newcommand{\mc}{\mathcal}
\newcommand{\mb}{\mathbf}
\newcommand{\mbb}{\mathbb}
\newcommand{\tr}{\mathrm{Tr}}

\newcommand{\dia}[1]{ \frac{1}{2}\left\| #1 \right\|_{\diamond} }
\newtheorem{theorem}{Theorem}

\newtheorem{lemma}{Lemma}

\newtheorem{result}{Result}

\newtheorem{corollary}{Corollary}

\newcommand{\D}{\mathcal{D}}

\newcommand{\M}{\mathcal{M}}

\newcommand{\F}{\mathbb{F}}

\newcommand{\R}{\mathcal{R}}

\graphicspath{{./figure/}}

\begin{document}
\title{One-shot dynamical resource theory}
\begin{abstract}
A fundamental problem in resource theory is to study the manipulation of the resource. Focusing on a general dynamical resource theory of quantum channels, here we consider tasks of one-shot resource distillation and dilution with a single copy of the resource. For any target of unitary channel or pure state preparation channel, we establish a universal strategy to determine upper and lower bounds on rates that convert between any given resource and the target. 
We show that the rates are related to resource measures based on the channel robustness and the channel hypothesis testing entropy, with regularization factors of the target resource measures. 
The strategy becomes optimal with converged bounds when the channel robustness is finite and measures of the target resource collapse to the same value. 
The single-shot result also applies to asymptotic parallel manipulation of channels to obtain asymptotic resource conversion rates. 
We give several examples of dynamical resources, including the purity, classical capacity, quantum capacity, non-uniformity, coherence, and entanglement of quantum channels.
Our results are applicable to general dynamical resource theories with potential applications in quantum communication, fault-tolerant quantum computing, and quantum thermodynamics. 


\end{abstract}
\author{Xiao Yuan}
\email{xiaoyuan@pku.edu.cn}
\affiliation{Center on Frontiers of Computing Studies, Department of Computer Science, Peking University, Beijing 100871, China}
\affiliation{Stanford Institute for Theoretical Physics, Stanford University, Stanford California 94305, USA}

\author{Pei Zeng}
\email{qubitpei@gmail.com}

\author{Minbo Gao}
\affiliation{Center for Quantum Information, Institute for Interdisciplinary Information Sciences, Tsinghua University, Beijing 100084, China}

\author{Qi Zhao}
\affiliation{Joint Center for Quantum Information and Computer Science, University of Maryland, College Park, Maryland 20742, USA}

\maketitle

\section{Introduction}
Quantum resource theories have played important roles in understanding quantum information and quantum computing tasks~\cite{horodecki2013quantumness,coecke_2016,chitambar2018quantum}. 
For example, the theories of entanglement~\cite{PhysRevA.53.2046,PhysRevLett.78.2275,RevModPhys.81.865} and magic~\cite{veitch2012negative,veitch2014resource,PhysRevLett.118.090501} have enabled us to quantitatively measure them and study the capability and limit of quantum communication and fault-tolerant quantum computing protocols.
Most earlier works have focused on static resources of quantum states and their manipulation with quantum channels. Regarding quantum channels themselves as dynamical objects, many recent works have been studying various properties of dynamical processes and their application in different tasks~\cite{quantifying17sr,Diaz2018usingreusing,dynamicalCoherence,theurer2018quantifying,PhysRevLett.125.130401,Rosset2018resource,yuan2019robustness,takagi_2020,2018arXiv181110307K,zhuang2018resource,seddon2019quantifying, wang2019quantifying,takagi2019general,PhysRevLett.122.140403, liu2019operational,liu2019resource,2020arXiv201011822F,2020arXiv201011942R,PhysRevResearch.1.033169,takagi2020optimal,bauml2019resource,dynamicalEntanglement,PhysRevLett.125.040502}. 
With the help of superchannels~\cite{chiribella2008transform,gour2019comparison}, which transforms a channel to a channel, these existing works are also unified under the resource theory language.


A general resource theory provides a consistent framework with monotones characterizing the resource object content, and studies central tasks such as resource conversion and their quantification. Specifically, the framework defines the so-called free resource (objects with zero-resource) and free operations (transformations that can be freely applied), and the task asks how a resource object could be converted to one another. 
The framework and resource monotones for dynamical resources have been studied for specific channel properties, such as classical and quantum channel capacities~\cite{Rosset2018resource,2018arXiv181110307K,yuan2019robustness,takagi_2020}, dynamical entanglement~\cite{bauml2019resource,dynamicalEntanglement,PhysRevLett.125.040502}, channel coherence~\cite{Mani2015cohering,Zanardi2017measures,Zanardi2017coherence,BU20171670,Dana2018resource,dynamicalCoherence,Diaz2018usingreusing,PhysRevLett.125.130401,theurer2018quantifying}, etc.  
While, since different dynamical resources generally have distinct structures, whether they could be studied in a more coherent and unified approach has been under active investigation~\cite{PhysRevA.101.062315,PhysRevLett.125.060405, liu2019operational,liu2019resource,2020arXiv201011942R,2020arXiv201011822F}. 

A seminar achievement along this direction is the unified theory for one-shot state distillation and dilution~\cite{2019arXiv190405840L}. However, dynamical resource theories, in general, have richer contents with more complicated channel transformations~\cite{chiribella2008transform,gour2019comparison} and monotones~\cite{PhysRevLett.123.150401}, as well as many diverse classes of resources that describe different ways of processing information and manipulating static state resources.
For example, considering how a channel processes the input information, e.g., whether any information or any quantum information is preserved, it corresponds to intrinsic channel resources, e.g., classical channel capacity~\cite{takagi_2020} and quantum channel capacity~\cite{Rosset2018resource,yuan2019robustness}, which are beyond state resource theories. Therefore, whether the unified theory for quantum states~\cite{2019arXiv190405840L} apply to dynamical quantum processes remains open.

In this work, we consider a general resource theory of dynamical processes of quantum channels and study their conversion under free superchannels. Focusing on a target unitary $\mc U$ or a pure state preparation channel $\mc G$, we investigate the distillation and dilution processes that convert between any channel $\mc N$ and several copies of the target. We show that the distillation or dilution rate, i.e., the maximal or minimal number of the target channel, is related to resource measures of $\mc N$ based on robustness and hypothesis testing entropy, as well as regularized quantifiers of the target channel. Our result becomes optimal when the robustness measure is finite and the target quantifiers collapse. We consider several example channel resources, describing intrinsic channel properties of purity, classical capacity, and quantum capacity, and state-induced channel properties of non-uniformity, coherence, and entanglement. We apply our results to obtain nearly optimal transformations of those dynamical resources. 

\section{Framework}
\subsection{Dynamical resource theories}
We first consider a general framework for resource $T$, which consists of the set of free resource objects $\mbb F_T$ and the set of free operations $\mbb O_T$. The two sets are related by the \emph{golden rule}, which requires that for any operation in $\mbb O_T$, it can only map a free resource object to a free one. 
For static resource theories, the objects are quantum states and the operations are quantum channels. 
For example, considering the resource theory of coherence~\cite{aberg2006quantifying,Baumgratz14,RevModPhys.89.041003}, the set of free states $\mbb{F}_{Coh}$, also known as incoherent states, are $\sigma=\sum_ip_i\ket{i}\bra{i}$, which are diagonal in the computational basis $\{\ket{i}\}$. For the set of free operations $\mbb{O}_T$, there are several different definitions, such as the maximally incoherent operation (MIO)~\cite{aberg2006quantifying}, dephasing-covariant incoherent operation~\cite{Chitambar16prl,Marvian16}, the incoherent operation~\cite{Baumgratz14}, etc, which could lead to different resource manipulation capabilities~\cite{Winter16,zhao2018oneshot,zhao2018oneshotdistill,regula2018one}. The largest set of free operations allowed by the golden rule is called the resource non-generating operations (RNG), as $\{\Phi(\sigma)\in\mbb{F}_T,\,\forall \sigma\in\mbb{F}_T\}$, e.g., the MIO for coherence. Any other definition of free operations is a subset of RNG operations. 

For a dynamical resource, the objects are quantum channels and the operations are superchannels~\cite{chiribella2008transform,gour2019comparison}. Specifically, for a system $A$, we denote its Hilbert space as $\mbb{H}(A)$ and linear operators as $\mbb{L}(A)$. A quantum channel $\mc{N}^{A\rightarrow B}$ from system $A$ to system $B$ is a completely positive (CP) and trace preserving (TP) linear map from $\mbb{L}(A)$ to $\mbb{L}(B)$. Denote the set of linear maps as $\mbb{L}^{A\rightarrow B}$. Regarding a channel as a dynamical object, a quantum superchannel $\mb \Lambda^{AB\rightarrow CD}$ from $\mc N^{A\rightarrow B}$ to $\mc M^{C\rightarrow D}$ is a supermap from $\mbb L^{A\rightarrow B}$ to $\mbb L^{C\rightarrow D}$. A supermap is a superchannel when it is completely CP preserving and TP preserving~\cite{gour2019comparison}. We refer to Appendix
for the exact definition of superchannels and their properties. Here we only need to notice that a superchannel can always be written as  
\begin{equation}
    \mb{\Lambda}(\mc{N}^{A\to B}) = \mc{U}^{BE\to D} \circ \left( \mc{N}^{A\to B} \otimes \mc{I}^{E} \right) \circ \mc{V}^{C\to AE},
\end{equation}
where $E$ is an ancillary system and $\mc{U}$ and  $\mc{V}$ are quantum channels.

An example dynamical resource is quantum memory or channel capacity~\cite{Rosset2018resource,yuan2019robustness}, which measures the capability of the channel in preserving quantum information or quantum entanglement. The set of free resources $\mbb F_{QC}$ are measure and prepare channels $\mc M(\rho)=\sum_i \tr[\rho O_i]\sigma_i$ with positive observable value measure $\{O_i\}$ and quantum states $\{\sigma_i\}$. Free operations could be RNG superchannels that map free channels to free channels~\cite{yuan2019robustness} or physically realizable ones as introduced in Ref.~\cite{Rosset2018resource}. Unitary channels are regarded as the optimal resource since they fully preserve the input information.

Meanwhile, for any state resource theory, one can define a corresponding channel resource theory to characterize its capability in manipulating the state resource~\cite{liu2019operational,liu2019resource,theurer2018quantifying}. Focusing on the capability of creating the resource, one can regard the set of RNG channels (or its subset) as a free dynamical resource, which cannot create a static resource from a non-resourceful state. Free operations on the dynamical resource could be defined as the set of RNG superchannels or its subset such as $\mb{\Lambda}(\mc{N}^{A\to B}) = \mc{M}^{BE\to D} \circ \left( \mc{N}^{A\to B} \otimes \mc{I}^{E} \right) \circ \mc{N}^{C\to AE}$ with RNG channels $\mc M$ and $\mc N$. One can regard a unitary that rotates a free state to the maximal resource as the target channel. Considering the interplay between state and channel resource theories, one can also choose the deterministic maximal resource state generation channel as the target.

\subsection{Resource monotones}
Based on the framework, now we introduce monotones to characterize the resource. We focus on quantum channels and consider a function $\mu$ that maps a channel object to a real number. It is called a resource monotone if it cannot be increased under free operations, i.e.,  $\mu(\mc N) \ge \mu(\mb\Lambda(\mc N)), \,\forall\mb\Lambda\in\mbb O$. We first introduce the monotone based on how robust the object is against the noise of free resource,
\begin{equation}
\begin{aligned}
 		R_T(\mc N) &= \min_{\mc M\in\F_T,}\left\{\lambda\ge 0:  \frac{1}{1+\lambda}\mc N+\frac{\lambda}{1+\lambda}\mc M\in\F_T\right\},\\
\end{aligned}
\end{equation}
which is monotonically non-increasing under any RNG operations. It is called robustness or free robustness since the optimization is over free resources. The measure $R_T(\mc N)$ could be infinite when the set of free resource does not span the whole space. {For example, when the robustness of state resources is infinite, the state-induced channel resource also have infinite robustness.} In this case, we consider the general robustness 
\begin{equation}
\begin{aligned}
		R_{G,T}(\mc N) &= \min_{\mc M}\left\{\lambda\ge 0: \frac{1}{1+\lambda}\mc N+\frac{\lambda}{1+\lambda}\mc M\in\F_T\right\},\\
\end{aligned}
\end{equation}
where the channel $\mc M$ could be an arbitrary CPTP map. We note that by defining the Choi state $\Phi^+_{\mc N} = \mc I\otimes \mc N(\Phi^+)$ with the maximally entangled state $\Phi^+$, the above optimization over channel can be re-expressed as an optimization over Choi states or any bipartite state with full Schmidt rank. 

In the following, we will show that the robustness or the generalized robustness will play a key role in characterizing one-shot resource dilution. To have a even clearer connection, we consider the log versions of the two measures as
\begin{equation}
    \begin{aligned}
    LR_T(\mc N) &= \log(1+R_T(\mc N)),\\ 
    D_{\max,T}(\mc N) &= \log(1+R_{G,T}(\mc N)).
    \end{aligned}
\end{equation}
Here we denote the logarithm version of the generalized robustness to be  $D_{\max,T}$ since it coincides with the definition based on the max-relative entropy of channels. 
Furthermore, for any resource monotone, we can smooth it by allowing a nonzero error parameter $\epsilon$. For example, the smoothed logarithmic robustness and max-entropy measures are
\begin{equation}\label{Eq:smoothLR}
\begin{aligned}
LR_T^\epsilon(\mc N)&= \min_{\dia{\mc{N}' - \mc{N}}\leq \epsilon }LR_T(\mc N'),\\
    D^\epsilon_{\max,T}(\mc N) &= \min_{\dia{\mc{N}' - \mc{N}}\leq \epsilon } D_{\max,T}(\mc{N}'),
\end{aligned}
\end{equation}
where $\dia{\mc{N}' - \mc{N}} = \frac{1}{2}\max_{\psi} \| (\mc{N'}\otimes\mc{I})(\psi) - (\mc{N}\otimes\mc{I})(\psi) \|_1$ is the diamond norm of channels with optimization over all states from the channel input and any ancillary input.

To characterize the one-shot distillation process, we consider another type of resource monotone based on the channel hypothesis testing entropy,
\begin{equation} \label{eq:DHeps}
\begin{aligned}
D_{H}^\epsilon(\mc{N}||\mc{M}) &= \max_{\psi} D_{H}^\epsilon(\mc{N}\otimes \mc I(\psi)||\mc{M}\otimes \mc I(\psi)) , \\
\end{aligned}
\end{equation}
with the state hypothesis testing entropy
\begin{equation}
	D_{H}^\epsilon(\rho\|\sigma) = \max_{\substack{0 \leq A \leq I \\ \tr[A\rho] \geq 1 - \epsilon }} -\log \tr[A\sigma].
\end{equation}
We note that the smooth is called operator smoothing, which is different from Eq.~\eqref{Eq:smoothLR}.
The induced resource monotone is defined by minimizing $D_{H}^\epsilon(\mc{N}||\mc{M})$ over all free resources as 
\begin{equation}
	D_{H,T}^\epsilon(\mc{N}) = \min_{\mc{M}\in \mbb{F}_T} D_H^\epsilon(\mc{N}||\mc{M}).
\end{equation}
Operationally, the measure $D_{H,T}^\epsilon(\mc{N})$ characterizes asymmetric discrimination between $\mc N$ and any free resource. 

We prove in Appendix
that all the above measures satisfy the monotonicity requirement and study their other properties such as convexity.  

\begin{lemma}
The quantifiers $D^\epsilon_{\max,T}(\mc N)$, $LR^\epsilon_T(\mc N)$, and $D_{H,T}^\epsilon(\mc{N})$ are resource monotones. 
\end{lemma}

The channel hypothesis testing entropy $D_{H}^\epsilon(\mc{N}||\mc{M})$ in Eq.~\eqref{eq:DHeps} involves an optimization over all input states that are possibly entangled with any ancillary system. Considering the maximally entangled state or states without entanglement as the input, we define other entropy measures
\begin{equation} \label{eq:}
\begin{aligned}
D_{\tilde H}^\epsilon(\mc{N}||\mc{M}) &= D_H^\epsilon(\Phi^+_{\mc N}||\Phi^+_{\mc M}), \\
    D_{\hat{H}}^\epsilon(\mc{N}||\mc{M})&= \max_{\psi} D_{H}^\epsilon(\mc{N}(\psi)||\mc{M}(\psi)).
\end{aligned}
\end{equation}
We note that 
\begin{equation}
    D_{H}^\epsilon(\mc{N}||\mc{M})\ge \max\{D_{\tilde H}^\epsilon(\mc{N}||\mc{M}), D_{\hat{H}}^\epsilon(\mc{N}||\mc{M})\}, 
\end{equation}
yet the order between $D_{\tilde{H}}^\epsilon(\mc{N}||\mc{M})$ and $D_{\hat{H}}^\epsilon(\mc{N}||\mc{M})$ is case dependent~\cite{yuan2019hypothesis}. We also define the corresponding resource quantifiers as
\begin{equation} \label{eq:}
\begin{aligned}
	D_{\tilde{H},T}^\epsilon(\mc{N}) &= \min_{\mc{M}\in \mbb{F}_T} D_{\tilde H}^\epsilon(\mc{N}||\mc{M}),\\
    D_{\hat{H},T}^\epsilon(\mc{N}) &= \min_{\mc{M}\in \mbb{F}_T} D_{\hat H}^\epsilon(\mc{N}||\mc{M}),\\	
\end{aligned}
\end{equation}
which will be useful for bounding the achievability part of one-shot resource distillation.




\subsection{Collapse of resource quantifiers}
The five resource quantifiers are not independent and they are ordered as
\begin{equation}
\begin{aligned}
    LR_{T}^\epsilon(\mc N)  &\ge D_{\max,T}^\epsilon(\mc N)
    \ge D_{{H},T}^\epsilon(\mc N)\ge  ... \\
    ...&\ge \{D_{\tilde{H},T}^\epsilon(\mc N), D_{\hat{H},T}^\epsilon(\mc N)\} ,
\end{aligned}    
\end{equation}
and whether the equal signs could  be achieved is now under investigation.
We first consider a state-induced channel resource with the target being a pure state preparation channel $\mc G(\rho)=\psi$ with pure state $\psi$. 
We show that the four resource monotones of $\mc G$ with $\epsilon=0$ are bounded by monotones of the state resource.

\begin{lemma}\label{lamma:collapseineq}
Consider a state induced channel resource with a state preparation channel, $\mc G(\rho)=\psi$, $\forall \rho$, with pure state $\psi$.
Suppose that free channels are $RNG$ channels of the state resource theory and the maximally mixed state is a free resource state, we have 
\begin{equation}
\begin{aligned}
   D^0_{\max, T}(\psi) &\ge D^0_{\max, T}(\mc G)\ge  D_{H, T}^0(\mc G) \ge ...\\ ... &\ge \{D_{\tilde{H},T}^0(\mc G), D_{\hat{H},T}^0(\mc G)\}\ge D_{ H, T}^0(\psi),
\end{aligned}
\end{equation}
where $D^0_{\max, T}(\psi)$ and $D_{ H, T}^0(\psi)$ are state resource monotones with respect to the set of free state resource.
\end{lemma}

\noindent It has been shown that the monotones of the maximal resource, i.e., the golden state, collapse to the same value when the set of free states is a convex hull of pure free states~\cite{2019arXiv190405840L}. Therefore the resource monotones of the state preparation channel also collapse.
\begin{lemma}\label{lemma:collapse}
For a state resource theory, suppose the set of free states is a convex hull of pure  states and includes all maximally mixed states. Suppose $\psi$ is the golden state resource that maximize $D_{ H, T}^0(\psi)$. 
Consider the corresponding channel resource with $RNG$ free channels. The resource monotones of the state preparation channel, $\mc G(\rho)=\psi$, $\forall \rho$, collapse
\begin{equation}
\begin{aligned}
   D^0_{\max, T}(\mc G)= D_{H, T}^0(\mc G)= D_{\tilde H, G}^0(\mc G) =D_{\hat H, G}^0(\mc G)=D_{ H, T}^0(\psi).
\end{aligned}
\end{equation}
\end{lemma}

\noindent Note that even though the logarithmic robustness is not included in the theorem, it also coincides with the measures (when it is finite) for resources such as dynamical entanglement as we show in Sec.~\ref{Sec:examples}. 


When $\mc N$ is a unitary channel $\mc U$,  the hypothesis testing entropy monotones for $\epsilon=0$ equal
\begin{equation}
\begin{aligned}
    D_{H,T}^0(\mc{U}) &= -\log F_T(\mc U),\\
    D_{\tilde{H},T}^0(\mc{U}) &= -\log \tilde{F}_T(\mc U),
\end{aligned}
\end{equation}
which are closely related to channel fidelity measures 
\begin{equation}\label{Eq:psioptim}
\begin{aligned}
	 F_T(\mc N) &= \max_{\mc M\in\mbb{F}_T}  \min_{\psi} F(\mc N(\psi)\otimes \mc I, \mc M(\psi)\otimes \mc{I} ),\\
	 \tilde{F}_T(\mc N) &= \max_{\mc M\in\mbb{F}_T}  F(\mc N(\Phi^+)\otimes\mc{I},\mc M(\Phi^+)\otimes \mc{I}),
\end{aligned}	 
\end{equation}
and state fidelity $F(\rho,\sigma)=(\tr[\sqrt{\sqrt{\rho}\sigma\sqrt{\rho}}])^2$. It is an open problem when the fidelity measures will be the same as the robustness measures. Nevertheless, for each resource theory, we can explicitly lower bound $D_{\tilde{H},T}^0(\mc{U})$ via the Choi state of $\mc U$ and a specific free resource, and upper bound $LR(\mc U)$ or $D_{\max}(\mc U)$ via an explicit decomposition. We can then prove the collapse of the measures by showing the convergence of the bounds, as we discuss in Sec.~\ref{Sec:examples}.

We also consider regularized measures of $n$ copies of resources with zero smooth as
\begin{equation}
\begin{aligned}
    {{m}_{LR,T}(\mc{N}^{\otimes n})} &=  LR^0_{T}(\mc{N}^{\otimes n})/n,\\
    {{m}_{\max,T}(\mc{N}^{\otimes n})} &=  D^0_{\max,T}(\mc{N}^{\otimes n})/n,\\
    m_{H,T}(\mc{N}^{\otimes n}) &= D_{H,T}^0(\mc{N}^{\otimes n})/n,\\
    m_{\tilde{H},T}(\mc N^{\otimes n}) &= D_{\tilde{H},T}^0(\mc N^{\otimes n})/n,\\
    m_{\hat{H},T}(\mc N^{\otimes n}) &= D_{\hat{H},T}^0(\mc N^{\otimes n})/n.
\end{aligned}
\end{equation}
We will show that different regularized measures of the target resource (being either $\mc G$ or $\mc U$) will be different normalizers to the upper and lower bounds of the distillation and dilution rates, as we shortly define.
The collapse of these measures will be crucial for exact distillation and dilution protocol. 






\section{One-shot distillation and dilution}
Now we consider general one-shot distillation and dilution between a given channel $\mc N$ and a target channel. We consider two types of targets with unitary channels and state preparation channels.

\subsection{Target --- unitary channels}
We first regard a unitary channel $\mc U$ as the target resource, and consider one-shot conversion between single use of a given resource $\mc N$ and multiple copies of $\mc U$. Specifically, we consider the maximal number of $\mc U$ that can be distilled from $\mc N$ and the minimal number of $\mc U$ that is required to synthesize $\mc N$, i.e., the tasks of resource distillation and dilution.  By allowing a smoothing parameter $\epsilon\in[0,1]$, we define the distillation and dilution rates as
\begin{equation}\label{eq:distilldilution}
\begin{aligned}
\R^{1,\epsilon}_{d,T}(\mc{N}) &= \max_{\mb{\Lambda} \in \mc{O}} \left\{ n:  \dia{ \mb{ \Lambda(\mc{N}) - \mc{U}^{\otimes n} } } \leq \epsilon \right\},\\
\R^{1,\epsilon}_{c,T}(\mc{N}) &= \min_{\mb{\Lambda} \in \mc{O}} \left\{ n: \dia{ \mb{ \Lambda(\mc{U}^{\otimes n}) - \mc{N} } } \leq \epsilon \right\}.
\end{aligned}
\end{equation}
Here the optimization is over all possible free operations and the parameter $\epsilon$ allows a certain error in the process. Determining the distillation and dilution rates is a central task in resource theories and it is in general challenging and resource-dependent for an arbitrary set of $\mc O$. However, when focusing on the maximal set of free operations, i.e., RNG superchannels, we can provide a unified approach to bound the distillation and dilution rates separately. Depending on whether the robustness measure is finite, we have two general schemes that are based on the optimization of the robustness and the generalized robustness measures. 

\subsubsection{Finite robustness}
We first show the one-shot distillation and dilution results given finite robustness measure of the channel $\mc N$. 

\begin{theorem} \label{thm:unitary_finite_distill}
When the free robustness measure $R(\mc{N})$ is finite for all the channels. For any given channel $\mc{N}$, the one-shot distillation rate $\R^{1,\epsilon}_{d,T}$ of the target resource $\mc{U}$ is given by
\begin{equation}
n_l\le \R^{1,\epsilon}_{d,T}(\mc{N}) \le D^{\epsilon}_{H,T} (\mc{N})/m_{H,T}(\mc U^{\otimes n_u}) 
\end{equation}
with $n_l=\max\{n: n\le D_{H,T}^\epsilon(\mc{N})/{{m}_{LR,T}(\mc{U}^{\otimes n})}\}$ and $n_u=\max\{n: n\le D^\epsilon_{H,T}(\mc{N})/m_{H,T}(\mc U^{\otimes n}) \}$.  
\end{theorem}

\begin{theorem} \label{thm:unitary_finite_dilute}
When the free robustness measure $R(\mc{N})$ is finite for all the channels. For any given channel $\mc{N}$, the one-shot dilution rate $\R^{1,\epsilon}_{c,T}$ of the target resource $\mc{U}$ is given by
\begin{equation}
LR^{\epsilon} (\mc{N})/{{m}_{LR,T}(\mc{U}^{\otimes n_l})}\le \R^{1,\epsilon}_{c,T}(\mc{N}) \le n_u,
\end{equation}
with $n_l = \min\{n\in \mbb D:n\ge LR^\epsilon_T(\mc{N})/{{m}_{LR,T}(\mc{U}^{\otimes n})} \}$ and $n_u = \min\{n\in \mbb D: n\ge LR^\epsilon_T(\mc{N})/m_{H,T}(\mc U^{\otimes n})\}$. 
\end{theorem}


The above results become clearer when the regularized measures for $\mc U$ collapse. In this case, we can use resource monotones to exactly characterize the one-shot distillation and dilution rates.
\begin{corollary} \label{cor:unitary_finite_distill_equal}
When the robustness measure is finite and ${{m}_{LR,T}(\mc{G}^{\otimes n})} =  m_{H,T}(\mc{G}^{\otimes n}) = c$ for all $n$, we have
\begin{equation} 
	\lfloor D^{\epsilon}_{H,T} (\mc{N})/c \rfloor \le \R^{1,\epsilon}_{d,T}(\mc{N}) \le  D^{\epsilon}_{H,T} (\mc{N})/c. 
\end{equation}

\end{corollary}

\begin{corollary} \label{cor:unitary_finite_dilute_equal}
When the robustness measure is finite and ${{m}_{LR,T}(\mc{U}^{\otimes n})} = m_{H,T}(\mc U^{\otimes n}) = c$ for all $n$, we have
\begin{equation}
	LR^{\epsilon}_{T} (\mc{N})/c\le \R^{1,\epsilon}_{c,T}(\mc{N}) \le \lceil LR^{\epsilon}_{T} (\mc{N})/c \rceil. 
\end{equation}

\end{corollary}
\noindent We will show that the collapse of measures for unitary channels indeed happens in the quantum capacity and dynamical entanglement resource theories.

\subsubsection{Infinite robustness}
When a channel $\mc N$ has infinite free robustness, we can instead construct one-shot distillation and dilution processes based on the generalized robustness measure. 

\begin{theorem} \label{thm:unitary_infinite_distill}
When there exist quantum channels with infinite free robustness, for a given channel $\mc{N}$, the one-shot distillation rate is alternatively upper bounded by
\begin{equation}
\R^{1,\epsilon}_{d,T}(\mc{N}) \le D^{\epsilon}_{H,T} (\mc{N})/m_{H,T}(\mc U^{\otimes n_u}) 
\end{equation}
where $n_u=\max\{n: n\le D^\epsilon_{H,T}(\mc{N})/m_{H,T}(\mc U^{\otimes n}) \}$. 
\end{theorem}

\begin{theorem} \label{thm:unitary_infinite_dilute}
When there exist quantum channels with infinite free robustness, for a given channel $\mc{N}$, the one-shot dilution rate is lower bounded by
\begin{equation}
\R^{1,\epsilon}_{c,T}(\mc{N}) \ge D_{\max,T}^{\epsilon} (\mc{N})/{{m}_{\max,T}(\mc{U}^{\otimes n_l})}  ,
\end{equation}
Moreover, if the target channel satisfies the constant trace condition, i.e, $\tr[\Phi^{CD}_{\mc{C}} \Phi^{CD}_{\mc U^{\otimes n}}]$ is a constant for all free channels $\mc{C}$ and a fixed $n$,
we have
\begin{equation}
 \R^{1,\epsilon}_{c,T}(\mc{N}) \le n_u.
\end{equation}
with $n_l = \min\{n:n \ge D_{\max,T}^\epsilon(\mc{N})/{m}_{\max,T}(\mc{U}^{\otimes n}) \}$ and $n_u=\min\{n\in \mbb D: n\ge D_{\max,T}^\epsilon(\mc{N})/m_{\tilde H,T}(\mc U^{\otimes n})\}$.
\end{theorem}
We note that the major difference to the results with finite robustness is that the achievability part is replaced with $D_{\tilde H,T}^\epsilon$ or $m_{\tilde H}(\mc U^{\otimes n})$. This is because here the distillation and dilution superchannels are constructed based on inputting the maximally entangled state,  whereas the superchannels used in the previous part does not have this assumption. 
When the regularized measures collapse, we have bounds as follows. 

\begin{corollary}\label{cor:unitary_infinite_dilute_equal}
When the robustness measure is infinite and ${{m}_{\max,T}(\mc{U}^{\otimes n})} = m_{\tilde H,T}(\mc U^{\otimes n}) = c$, we have
\begin{equation} 
	D_{\max,T}^{\epsilon} (\mc{N})/c\le \R^{1,\epsilon}_{c,T}(\mc{N}) \le \lceil D_{\max,T}^{\epsilon} (\mc{N})/c \rceil. 
\end{equation}
\end{corollary}

\subsection{Target --- state preparation channels}
Now consider that the target channel is a pure state preparation or resource generation channel $\mc{G}$, which maps any input state to a pure output state. We can similarly construct the distillation and dilution channels. The result also depends on whether the robustness is finite and the constructions assume the input states are maximally entangled states. We summarize the result with both finite and infinite robustness as follows.

\subsubsection{Finite robustness}

\begin{theorem} \label{thm:stateprep_finite_distill}
When the free robustness $R(\mc{N})$ is finite for all the channels, for a given $\mc{N}$, the one-shot distillation rate of the optimal state preparation channel $\mc{G}$ is bounded by
\begin{equation}
n_l\le \R^{1,\epsilon}_{d,T}(\mc{N}) \le D^{\epsilon}_{H,T} (\mc{N})/m_{\hat{H},T}(\mc{G}^{\otimes n_u}), 
\end{equation}
with $n_l=\max\{n: n \le D_{\tilde{H},T}^\epsilon(\mc{N}) / {{m}_{LR,T}(\mc{G}^{\otimes n})}\}$ and $n_u=\max\{n: n \le D^\epsilon_{H,T}(\mc{N})/m_{\hat{H},T}(\mc{G}^{\otimes n}) \}$. 
\end{theorem}

\begin{theorem} \label{thm:stateprep_finite_dilute} 
When the free robustness $R_T(\mc{N})$ is finite for all the channels, for a given channel $\mc{N}$, the one-shot dilution rate $\R^{1,\epsilon}_{c,T}(\mc{N})$ of the optimal state preparation channel $\mc{G}$ under the RNG operations is bounded by
\begin{equation}
LR^{\epsilon}_{T} (\mc{N})/{{m}_{LR,T}(\mc{G}^{\otimes n_l})}\le \R^{1,\epsilon}_{c,T}(\mc{N}) \le n_u,
\end{equation}
with $n_l = \min\{n\in \mbb D:n \ge LR_{T}^\epsilon(\mc{N})/{m}_{LR,T}(\mc{G}^{\otimes n}) \}$ and $n_u = \min\{n\in \mbb D: n\ge LR_T^\epsilon(\mc{N})/m_{\tilde{H},T}(\mc G^{\otimes n}) \}$.
\end{theorem}

When the regularized measures collapse, we have bounds as follows. 
\begin{corollary} \label{cor:stateprep_finite_distill_equal}
When the robustness measure is finite with  ${{m}_{LR,T}(\mc{G}^{\otimes n})} =  m_{\hat{H},T}(\mc G^{\otimes n}) = c$, we have
\begin{equation}
	\lfloor D^{\epsilon}_{\tilde H,T} (\mc{N})/c \rfloor \le \R^{1,\epsilon}_{d,T}(\mc{N}) \le D^{\epsilon}_{H,T} (\mc{N})/c. 
\end{equation}

\end{corollary}

\begin{corollary} \label{cor:stateprep_finite_dilute_equal}
When the robustness measure is finite and ${{m}_{LR,T}(\mc{G}^{\otimes n})} = m_{\tilde{H},T}(\mc G^{\otimes n}) = c$, we have
\begin{equation}
	LR^{\epsilon}_{T} (\mc{N})/c\le \R^{1,\epsilon}_{c,T}(\mc{N}) \le \lceil LR^{\epsilon}_{T} (\mc{N})/c \rceil. 
\end{equation}
\end{corollary}

\subsubsection{Infinite robustness}

Similarly, when the free robustness of a channel $\mc{N}$ is infinite, we can evaluate the one-shot distillation and dilution processes as follows.

\begin{theorem} \label{thm:stateprep_infinite_distill}
When there exist quantum channels with infinite free robustness, for a given channel $\mc{N}$, the one-shot distillation rate $\R^{1,\epsilon}_{d,T}(\mc{N})$ of the target state preparation channel is  bounded by
\begin{equation}
\R^{1,\epsilon}_{d,T}(\mc{N}) \le D^{\epsilon}_{H,T} (\mc{N})/m_{\hat{H},T}(\mc G^{\otimes n_u}) 
\end{equation}
where $ n_u=\max\{n: n \le D^\epsilon_{H,T}(\mc{N})/m_{\hat{H},T}(\mc G^{\otimes n}) \} $.
\end{theorem}

\begin{theorem} \label{thm:stateprep_infinite_dilute}
When there exist quantum channels with infinite free robustness, for a given channel $\mc{N}$, the one-shot dilution rate $\R^{1,\epsilon}_{c,T}(\mc{N})$ is bounded by
\begin{equation}
\R^{1,\epsilon}_{c,T}(\mc{N}) \ge D_{\max}^{\epsilon} (\mc{N})/{{m}_{\max}(\mc{G}^{\otimes n_l})}  ,
\end{equation}
Moreover, if the target channel satisfies the constant trace condition, i.e, $\tr[\Phi^{CD}_{\mc{C}} \Phi^{CD}_{\mc{G}^{\otimes n}}]$ is a constant for all free channels $\mc{C}$ and fixed $n$. Then we have:
\begin{equation}
 \R^{1,\epsilon}_{c,T}(\mc{N}) \le n_u.
\end{equation}
with $n_l = \min\{n:n\cdot{{m}_{\max}(\mc{}^{\otimes n})}\ge D_{\max}^\epsilon(\mc{N}) \}$ and $n_u=\min\{n\in \mbb D: n\ge D_{\max}^\epsilon(\mc{N})/m_{\tilde H}(\mc G^{\otimes n})\}$.
\end{theorem}

When the regularized measures collapse, we have bounds as follows. 

\begin{corollary} \label{cor:stateprep_infinite_dilute_equal}
When the robustness measure is infinite and ${{m}_{\max,T}(\mc{G}^{\otimes n})} = m_{\tilde H}(\mc G^{\otimes n}) = c$, we have
\begin{equation} 
	D_{\max}^{\epsilon} (\mc{N})/c \le \R^{1,\epsilon}_{c,T}(\mc{N}) \le \lceil D_{\max}^{\epsilon} (\mc{N})/c \rceil. 
\end{equation}
\end{corollary}

\section{Examples}\label{Sec:examples}

Here we consider several typical channel resource theories, describing intrinsic channel properties and state-induced manipulation capabilities. We show how the general results in the previous section could be applied to specific resource theories. 
\subsection{Intrinsic channel resources}
We first discuss three intrinsic channel resource properties in preserving quantum information without relying on state resource theories. 

\subsubsection{Purity}
Almost all existing quantum properties relies on pure quantum states. Therefore, we define the \emph{purity} of a channel by its ability in producing pure states. Since the least pure state is the maximally mixed state  $\pi_d = I_d/d$ with dimension $d$, the least pure channel is the maximally depolarizing channels $\D(\rho) = \pi_d$. To quantify the purity of channels, we define the corresponding free resources to be the set of maximally depolarizing channels
\begin{equation}
    \F_{P} = \{\D | \D(\rho) = \pi_d\}. 
\end{equation}
Since an arbitrary channels could not be decomposed as linear combination of maximally depolarizing channels, 
the robustness is infinite. 

For the target channel, we consider it to be the identical qubit channel $\mc I_2$. We note that applying a unitary after a channel, i.e., $\bf \Lambda(\mc N)=\mc U\circ \mc N$ is a free superchannel, therefore it is equivalent to consider an arbitrary unitary as the target. 
Consider the qubit channel $\mc I_2$, we have
\begin{equation}
\begin{aligned}
    m_{\max,P}(\mc I_2^{\otimes n}) = m_{H,P}(\mc I_2^{\otimes n}) = m_{\tilde H,P}(\mc I_2^{\otimes n}) = 2\\
\end{aligned}
\end{equation}
for any integer $n$. Therefore we can apply Theorem~\ref{thm:unitary_infinite_distill} and Corollary~\ref{cor:unitary_infinite_dilute_equal} to have the one-shot distillation and dilution rates. We further prove refined bounds for the distillation part and we conclude our results as follows. 
\begin{result}
The one-shot distillation and dilution rates of channel purity under RNG superchannels are
\begin{equation}
    \begin{aligned}
    \lfloor D^{\epsilon}_{H,P} (\mc{N})/2 \rfloor \le \R^{1,\epsilon}_{d,P}(\mc{N}) \le  D^{\epsilon}_{H,P} (\mc{N})/2 ,\\
    D_{\max,P}^{\epsilon} (\mc{N})/2\le \R^{1,\epsilon}_{c,P}(\mc{N}) \le \lceil D_{\max,P}^{\epsilon} (\mc{N})/2 \rceil. 
    \end{aligned}
\end{equation}
\end{result}
We can apply the one-shot result to obtain the asymptotic resource conversion limit. Defining the asymptotic parallel distillation and dilution rates as
\begin{equation}
    \begin{aligned}
    \R^{\infty}_{d,T}(\mc{N}) &= \lim_{\epsilon\rightarrow0^+}\lim_{n\rightarrow\infty}\frac{1}{n}\R^{1,\epsilon}_{d,T}(\mc{N}^{\otimes n}),\\
\R^{\infty}_{c,T}(\mc{N}) &= \lim_{\epsilon\rightarrow0^+}\lim_{n\rightarrow\infty}\frac{1}{n}\R^{1,\epsilon}_{c,T}(\mc{N}^{\otimes n}).
    \end{aligned}
\end{equation}
Consider the channel relative entropy
\begin{equation}
    S(\mc N\|\mc M) = \max_{\psi} S(\mc N(\psi)\|\mc M(\psi)),
\end{equation}
and the channel entropy
\begin{equation}
    S(\mc N) = S(\mc N\|\mc D).
\end{equation}
With the asymptotic equal partition of channel entropy~\cite{gour2018entropy}, we can show that the asymptotic parallel distillation and dilution rates converges to $S(\mc N)$. 
\begin{result}
The asymptotic parallel distillation and dilution rates of channel purity under RNG superchannels are
\begin{equation}
    \begin{aligned}
    \R^{\infty}_{d,P}(\mc{N}) = \R^{\infty}_{c,P}(\mc{N}) = S(\mc N)/2.
    \end{aligned}
\end{equation}
\end{result}
\noindent Here we only consider parallel strategies and generalizing the results to incorporate adaptive strategies is an interesting future work.

\subsubsection{Classical capacity}
In addition to producing pure states, a higher level resource of a channel is its capability in preserving the input state information. We note that a replacement channel $\R(\rho) = \sigma$ outputs a state that is independent of the input. Therefore, we define the set of free resources to be 
\begin{equation}
    \F_{CC} = \{\R | \R(\rho) = \sigma \}
\end{equation}
We call it the \emph{classical capacity} or \emph{reversibility} of channels and we also note that it corresponds to the \emph{classical capacity} of quantum Shannon theory in quantum communication~\cite{takagi_2020}. 

Consider the qubit channel $\mc I_2$ as the target resource, which maximally preserves the input information. The robustness measure is infinite since linear combinations of replacement channels are still replacement maps. We thus focus on the generalized robustness measure and show that 
\begin{equation}
\begin{aligned}
    m_{\max,CC}(\mc I_2^{\otimes n}) = m_{H,CC}(\mc I_2^{\otimes n}) = m_{\tilde H,CC}(\mc I_2^{\otimes n}) = 2.
\end{aligned}
\end{equation}
Applying Theorem~\ref{thm:unitary_infinite_distill} and Corollary~\ref{cor:unitary_infinite_dilute_equal}, we have the one-shot distillation and dilution rates. 
\begin{result}
The one-shot distillation and dilution rates of channel classical capacity under RNG superchannels are
\begin{equation}
    \begin{aligned}
   &\R^{1,\epsilon}_{d,CC}(\mc{N}) \le D^{\epsilon}_{H,CC} (\mc{N})/2 ,\\
    D_{\max,CC}^{\epsilon} &(\mc{N})/2\le \R^{1,\epsilon}_{c,T}(\mc{N}) \le \lceil D_{\max,CC}^{\epsilon} (\mc{N})/2 \rceil. 
    \end{aligned}
\end{equation}
\end{result}
We also get the asymptotic conversions rates by applying the results to the asymptotic limit. Define measures  
\begin{equation}
    \begin{aligned}
         S^\infty_{T}(\mc N) &= \lim_{n\rightarrow\infty} \min_{\mc M\in\mbb F_{T}}\frac{1}{n}S(\mc N^{\otimes n} \|\mc M),\\
         \tilde S^\infty_{T}(\mc N) &= \lim_{n\rightarrow\infty} \min_{\mc M\in\mbb F_{T}}\frac{1}{n}S(\Phi_{\mc N^+}^{\otimes n}\|\Phi_{\mc M^+}),
    \end{aligned}
\end{equation}
where $\tilde S^\infty_{T}(\mc N)$ only consider inputs with the maximally entangled state. 
We have the asymptotic distillation and dilution rates with the help of tools from Ref.~\cite{takagi_2020,Fang_2020,PhysRevResearch.1.033169}.
\begin{result}
The asymptotic parallel distillation and dilution rates of channel classical capacity under RNG superchannels are
\begin{equation}
    \begin{aligned}
        R^{\infty}_{d,CC}(\mc{N}) \le  S^\infty_{CC}(\mc N)/2 = R^{\infty}_{c,CC}(\mc{N}).
    \end{aligned}
\end{equation}
\end{result}
    

\subsubsection{Quantum capacity}
The classical capacity is defined with respect to preserving any input state information. We can also focus on the \emph{quantum capacity} or \emph{quantum memory} in preserving quantum information~\cite{Rosset2018resource,2018arXiv181110307K,yuan2019robustness}. We consider a measure-and-prepare channel  $\M(\rho) = \sum_i \tr[\rho M_i] \sigma_i$ with POVM $\{M_i\}$ with POVM $\{M_i\}$ and quantum states $\{\sigma_i\}$, which destroys the input state by the measurement and only forwards the classical measurement outcome. We define the set of free resources to be
\begin{equation}
    \F_{QC} = \{\M | \M(\rho) = \sum_i \tr[\rho M_i] \sigma_i\}.
\end{equation}
Since the entanglement between the input state and any other systems are destroyed by a measure-and-prepare channel, the quantum capacity also characterizes the ability in preserving  quantum correlation. 



Consider the qubit channel $\mc I_2$ as the target resource, which preserves any quantum correlation. The robustness measure is finite and we have
\begin{equation}
\begin{aligned}
    m_{LR,QC}(\mc I_2^{\otimes n}) = m_{H,QC}(\mc I_2^{\otimes n}) = m_{\tilde H,QC}(\mc I_2^{\otimes n}) = 1.
\end{aligned}
\end{equation}
Applying Corollary~\ref{cor:unitary_finite_distill_equal} and \ref{cor:unitary_finite_dilute_equal}, we have the one-shot distillation and dilution rates.
\begin{result}
The one-shot distillation and dilution rates of channel quantum capacity under RNG superchannels are
\begin{equation}
    \begin{aligned}
\lfloor    D^{\epsilon}_{H,QC} (\mc{N})\rfloor\le \R^{1,\epsilon}_{d,QC}(\mc{N}) \le D^{\epsilon}_{H,QC} (\mc{N}) ,\\
    LR_{QC}^{\epsilon} (\mc{N})\le \R^{1,\epsilon}_{c,QC}(\mc{N}) \le \lceil LR_{QC}^{\epsilon} (\mc{N}) \rceil. 
    \end{aligned}
\end{equation}
\end{result}

We have the asymptotic distillation rate with the help of tools from Ref.~\cite{PhysRevResearch.1.033169}.
\begin{result}
The asymptotic parallel distillation  rate of channel quantum capacity under RNG superchannels is
\begin{equation}
    \begin{aligned}
        R^{\infty}_{QC,P}(\mc{N}) =   S^\infty_{QC}(\mc N).
    \end{aligned}
\end{equation}
\end{result}
\noindent Whether the dilution rate converges to $S^\infty_{QC}(\mc N)$ is an open question.


\subsection{State induced channel resources}
Meanwhile, for any state resource theory, it also induces channel resources. For example, focusing on the capability of generating state resources, the set of resource non-generating operations for states (or their subsets) could be regarded as the set of free channel resource. We consider several examples related to state resource theories of non-uniformity, coherence, and entanglement. 

\subsubsection{non-uniformity}
Considering the state resource theory of non-uniformity~\cite{Gour_2015}, the only free resource is the maximally mixed states and RNG operations are unital channels $\M_U(\pi_d) = \pi_{d'}$. We consider the corresponding dynamical resource with free resource 
\begin{equation}
    \F_{NU} = \{\M | \M(\pi_d) = \pi_{d'}\},
\end{equation}
which   characterizes the channel capability in generating non-maximally mixed states from free resource states. 
We regard the pure qubit state generation channel $\mc G_2(\rho)=\ket{\psi}\bra{\psi}$ with qubit state $\ket{\psi}$ as the target resource. Since linear combinations of unital channels are still unital, the robustness could be infinite and we consider the generalized robustness measure. According to Lemma~\ref{lemma:collapse}, we have
\begin{equation}\label{eq:m-non-uniformity}
    m_{\max,NU}(\mc G_2) = m_{\tilde H,NU}(\mc G_2) = 1,
\end{equation}
and the one-shot distillation and dilution rates after applying Theorem~\ref{thm:stateprep_infinite_distill} and Corollary~\ref{cor:stateprep_infinite_dilute_equal}.
\begin{result}
The one-shot distillation and dilution rates of generating non-uniform states under RNG superchannels are
\begin{equation}
    \begin{aligned}
   &\R^{1,\epsilon}_{d,NU}(\mc{N}) \le D^{\epsilon}_{H, NU} (\mc{N}) ,\\
    D_{\max,NU}^{\epsilon}& (\mc{N})\le \R^{1,\epsilon}_{c,NU}(\mc{N}) \le \lceil D_{\max,NU}^{\epsilon} (\mc{N}) \rceil. 
    \end{aligned}
\end{equation}
\end{result}




\subsubsection{Coherence}
With a computational basis $\{\ket{i}\}$, coherence describes the superposition power in the basis~\cite{aberg2006quantifying,Baumgratz14,RevModPhys.89.041003}. Free state resources are incoherent states $\delta=\sum_i p_i\ket{i}\bra{i}$, which are diagonal states, and the RNG free operations are maximally incoherent operations (MIO), i.e.,
$\mc M(\delta) = \delta'$. 
The corresponding dynamical resource for coherence has free resource
\begin{equation}
    \F_{Coh}=\{\mc M|\mc M(\delta) = \delta'\}.
\end{equation}
We choose the Hadamard gate $U_{Had}$ as the target unitary $\mc U_{Coh}(\rho)=U_{Had}\rho U_{Had}^\dag$ because it maps incoherent states to maximally coherent states. 
Since linear combinations of MIO channels are still MIO, the robustness could be infinite and we consider the generalized robustness measure. We have
\begin{equation}\label{eq:m-Non-coherence}
    m_{\max,Coh}(\mc U_{Had}) = m_{H,Coh}(\mc U_{Had}) = m_{\tilde H, Coh}(\mc U_{Had}) = 1,
\end{equation}
and the the one-shot distillation and dilution rates after applying Theorem~\ref{thm:unitary_infinite_distill} and Corollary~\ref{cor:unitary_infinite_dilute_equal}.
\begin{result}
With target channel $\mc U_{Had}$, the one-shot distillation and dilution rates of coherence under RNG superchannels are
\begin{equation}
    \begin{aligned}
    &\R^{1,\epsilon}_{d,Coh}(\mc{N}) \le  D^{\epsilon}_{H, Coh} (\mc{N}) ,\\
    D_{\max,Coh}^{\epsilon} &(\mc{N})\le \R^{1,\epsilon}_{c,Coh}(\mc{N}) \le \lceil D_{\max,Coh}^{\epsilon} (\mc{N}) \rceil. 
    \end{aligned}
\end{equation}
\end{result}

We can also consider the maximally coherent state generation channel $\mc G_{+}$ as the target, i.e., $\mc G_{+}(\rho)=(\ket{0}+\ket{1})/\sqrt{2}$. According to Lemma~\ref{lemma:collapse}, we have
\begin{equation}\label{eq:}
    m_{\max,Coh}(\mc G_{+}) = m_{ H}(\mc G_{+}) = m_{\tilde H}(\mc G_{+}) = 1,
\end{equation}
and therefore the distillation and dilution rates with Theorem~\ref{thm:stateprep_infinite_distill} and  Corollary~\ref{cor:stateprep_infinite_dilute_equal}. 

\begin{result}
The one-shot distillation and dilution rates of generating the maximally coherent state under RNG superchannels are
\begin{equation}
    \begin{aligned}
   &\R^{1,\epsilon}_{d,Coh}(\mc{N}) \le D^{\epsilon}_{H, Coh} (\mc{N}) ,\\
    D_{\max,Coh}^{\epsilon} &(\mc{N})\le \R^{1,\epsilon}_{c,Coh}(\mc{N}) \le \lceil D_{\max,Coh}^{\epsilon} (\mc{N}) \rceil. 
    \end{aligned}
\end{equation}
\end{result}

\subsubsection{Entanglement}
Entanglement measures the non-local correlation between multiple systems~\cite{PhysRevA.53.2046,PhysRevLett.78.2275,RevModPhys.81.865}. Considering bipartite entanglement as an example, free state resources are separable states $\sigma_{SEP} =\sum_i p_i \rho_i^A\otimes \rho_i^B$ with density matrices $\rho_i^A$ and $\rho_i^B$ on systems $A$ and $B$, respectively. We consider free dynamical resource as the set of completely RNG free operations
\begin{equation}
    \F_{Ent}=\{\mc M|\mc I\otimes \mc M(\sigma_{SEP})=\sigma_{SEP}',\forall \sigma_{SEP}\}.
\end{equation}
We choose the CNOT gate $U_{CNOT}$ as the target unitary $\mc U_{CN}(\rho)=U_{CNOT}\rho U_{CNOT}^\dag$, which can generate maximally entangled states from separable states. The robustness measure is finite and we have that
\begin{equation}
\begin{aligned}
    m_{LR,Ent}(\mc U_{CN}^{\otimes n}) = m_{H,Ent}(\mc U_{CN}^{\otimes n}) = m_{\tilde H,Ent}(\mc U_{CN}^{\otimes n}) = 2.
\end{aligned}
\end{equation}
Applying 
 Corollary~\ref{cor:unitary_finite_distill_equal} and \ref{cor:unitary_finite_dilute_equal}, we have the one-shot distillation and dilution rates.
\begin{result}
With target channel $\mc U_{CN}$, the one-shot distillation and dilution rates of entanglement under RNG superchannels are
\begin{equation}
    \begin{aligned}
    \lfloor D^{\epsilon}_{\tilde H,Ent} (\mc{N})/2\rfloor \le \R^{1,\epsilon}_{d,Ent}(\mc{N}) \le D^{\epsilon}_{H, Ent} (\mc{N}) /2,\\
    LR_{Ent}^{\epsilon} (\mc{N})/2\le \R^{1,\epsilon}_{c,Ent}(\mc{N}) \le \lceil LR_{Ent}^{\epsilon} (\mc{N})/2 \rceil. 
    \end{aligned}
\end{equation}
\end{result}

Now consider the quantum channel $\mc G_{\Phi^+}$ that deterministically generates the maximally  entangled state $\ket{\Phi^+}=(\ket{00}+\ket{11})/\sqrt{2}$, we have
\begin{equation}
\begin{aligned}
    m_{LR,Ent}(\mc G_{\Phi^+}^{\otimes n}) = m_{H,Ent}(\mc G_{\Phi^+}^{\otimes n}) = m_{\tilde H,Ent}(\mc G_{\Phi^+}^{\otimes n}) = 1.
\end{aligned}
\end{equation}
According to Corollary~\ref{cor:stateprep_finite_distill_equal} and \ref{cor:stateprep_finite_dilute_equal}, the one-shot distillation and dilution rates are
\begin{result}
The one-shot distillation and dilution rates of generating the maximally coherent state under RNG superchannels are
\begin{equation}
    \begin{aligned}
   \lfloor D^{\epsilon}_{\tilde H,Ent} (\mc{N})\rfloor\le \R^{1,\epsilon}_{d,Ent}(\mc{N}) \le D^{\epsilon}_{H, Ent} (\mc{N}) ,\\
    LR_{Ent}^{\epsilon} (\mc{N})\le \R^{1,\epsilon}_{c,Ent}(\mc{N}) \le \lceil LR_{Ent}^{\epsilon} (\mc{N}) \rceil. 
    \end{aligned}
\end{equation}
\end{result}




\section{Summary}
In this work, we study one-shot distillation and dilution of general dynamical resource theories. For any unitary channel or any pure state preparation channel as the target resource, we give universal upper and lower bounds to the distillation and dilution rates for conversions between any channel and the target resource. The schemes become optimal for several example resources with specific targets. We expect our results to have wide applications in quantum information science from quantum communication to  quantum computing. 


%

\begin{acknowledgements}
X.Y acknowledges support from the Simons Foundation.
Q.Z. acknowledges the support by the Department of Defense through the Hartree Postdoctoral Fellowship at QuICS. P.Z. and M.G. acknowledge support from the National Natural Science Foundation of China Grants No.~11875173 and No.~11674193, the National Key Research and Development Program of China Grant No.~2019QY0702 and No.~2017YFA0303903, and the Zhongguancun Haihua Institute for Frontier Information Technology.
\end{acknowledgements}

\emph{Note added.---}~Recently we became aware of two closely related works that independently study one-shot channel resource theories. The one by Bartosz Regula1 and Ryuji Takagi considered a general theory~\cite{dynamicalBartosz} and the other one by Ho-Joon Kim, Soojoon Lee, Ludovico Lami, and Martin B. Plenio~\cite{dynamicalBartosz} focused on quantum entanglement. We thank the authors for sharing their manuscripts and coordinating concurrent submission.

X.~Y.~and P.~Z.~contributed equally to this work.

\bibliographystyle{apsrev4-1}

\begin{thebibliography}{57}%
\makeatletter
\providecommand \@ifxundefined [1]{%
 \@ifx{#1\undefined}
}%
\providecommand \@ifnum [1]{%
 \ifnum #1\expandafter \@firstoftwo
 \else \expandafter \@secondoftwo
 \fi
}%
\providecommand \@ifx [1]{%
 \ifx #1\expandafter \@firstoftwo
 \else \expandafter \@secondoftwo
 \fi
}%
\providecommand \natexlab [1]{#1}%
\providecommand \enquote  [1]{``#1''}%
\providecommand \bibnamefont  [1]{#1}%
\providecommand \bibfnamefont [1]{#1}%
\providecommand \citenamefont [1]{#1}%
\providecommand \href@noop [0]{\@secondoftwo}%
\providecommand \href [0]{\begingroup \@sanitize@url \@href}%
\providecommand \@href[1]{\@@startlink{#1}\@@href}%
\providecommand \@@href[1]{\endgroup#1\@@endlink}%
\providecommand \@sanitize@url [0]{\catcode `\\12\catcode `\$12\catcode
  `\&12\catcode `\#12\catcode `\^12\catcode `\_12\catcode `\%12\relax}%
\providecommand \@@startlink[1]{}%
\providecommand \@@endlink[0]{}%
\providecommand \url  [0]{\begingroup\@sanitize@url \@url }%
\providecommand \@url [1]{\endgroup\@href {#1}{\urlprefix }}%
\providecommand \urlprefix  [0]{URL }%
\providecommand \Eprint [0]{\href }%
\providecommand \doibase [0]{http://dx.doi.org/}%
\providecommand \selectlanguage [0]{\@gobble}%
\providecommand \bibinfo  [0]{\@secondoftwo}%
\providecommand \bibfield  [0]{\@secondoftwo}%
\providecommand \translation [1]{[#1]}%
\providecommand \BibitemOpen [0]{}%
\providecommand \bibitemStop [0]{}%
\providecommand \bibitemNoStop [0]{.\EOS\space}%
\providecommand \EOS [0]{\spacefactor3000\relax}%
\providecommand \BibitemShut  [1]{\csname bibitem#1\endcsname}%
\let\auto@bib@innerbib\@empty
\bibitem [{\citenamefont {Horodecki}\ and\ \citenamefont
  {Oppenheim}(2013)}]{horodecki2013quantumness}%
  \BibitemOpen
  \bibfield  {author} {\bibinfo {author} {\bibfnamefont {M.}~\bibnamefont
  {Horodecki}}\ and\ \bibinfo {author} {\bibfnamefont {J.}~\bibnamefont
  {Oppenheim}},\ }\href@noop {} {\bibfield  {journal} {\bibinfo  {journal}
  {International Journal of Modern Physics B}\ }\textbf {\bibinfo {volume}
  {27}},\ \bibinfo {pages} {1345019} (\bibinfo {year} {2013})}\BibitemShut
  {NoStop}%
\bibitem [{\citenamefont {Coecke}\ \emph {et~al.}(2016)\citenamefont {Coecke},
  \citenamefont {Fritz},\ and\ \citenamefont {Spekkens}}]{coecke_2016}%
  \BibitemOpen
  \bibfield  {author} {\bibinfo {author} {\bibfnamefont {B.}~\bibnamefont
  {Coecke}}, \bibinfo {author} {\bibfnamefont {T.}~\bibnamefont {Fritz}}, \
  and\ \bibinfo {author} {\bibfnamefont {R.~W.}\ \bibnamefont {Spekkens}},\
  }\href {\doibase 10.1016/j.ic.2016.02.008} {\bibfield  {journal} {\bibinfo
  {journal} {Inf. Comput.}\ }\textbf {\bibinfo {volume} {250}},\ \bibinfo
  {pages} {59} (\bibinfo {year} {2016})}\BibitemShut {NoStop}%
\bibitem [{\citenamefont {Chitambar}\ and\ \citenamefont
  {Gour}(2018)}]{chitambar2018quantum}%
  \BibitemOpen
  \bibfield  {author} {\bibinfo {author} {\bibfnamefont {E.}~\bibnamefont
  {Chitambar}}\ and\ \bibinfo {author} {\bibfnamefont {G.}~\bibnamefont
  {Gour}},\ }\href {https://arxiv.org/abs/1806.06107} {\bibfield  {journal}
  {\bibinfo  {journal} {arXiv preprint arXiv:1806.06107}\ } (\bibinfo {year}
  {2018})}\BibitemShut {NoStop}%
\bibitem [{\citenamefont {Bennett}\ \emph {et~al.}(1996)\citenamefont
  {Bennett}, \citenamefont {Bernstein}, \citenamefont {Popescu},\ and\
  \citenamefont {Schumacher}}]{PhysRevA.53.2046}%
  \BibitemOpen
  \bibfield  {author} {\bibinfo {author} {\bibfnamefont {C.~H.}\ \bibnamefont
  {Bennett}}, \bibinfo {author} {\bibfnamefont {H.~J.}\ \bibnamefont
  {Bernstein}}, \bibinfo {author} {\bibfnamefont {S.}~\bibnamefont {Popescu}},
  \ and\ \bibinfo {author} {\bibfnamefont {B.}~\bibnamefont {Schumacher}},\
  }\href {\doibase 10.1103/PhysRevA.53.2046} {\bibfield  {journal} {\bibinfo
  {journal} {Phys. Rev. A}\ }\textbf {\bibinfo {volume} {53}},\ \bibinfo
  {pages} {2046} (\bibinfo {year} {1996})}\BibitemShut {NoStop}%
\bibitem [{\citenamefont {Vedral}\ \emph {et~al.}(1997)\citenamefont {Vedral},
  \citenamefont {Plenio}, \citenamefont {Rippin},\ and\ \citenamefont
  {Knight}}]{PhysRevLett.78.2275}%
  \BibitemOpen
  \bibfield  {author} {\bibinfo {author} {\bibfnamefont {V.}~\bibnamefont
  {Vedral}}, \bibinfo {author} {\bibfnamefont {M.~B.}\ \bibnamefont {Plenio}},
  \bibinfo {author} {\bibfnamefont {M.~A.}\ \bibnamefont {Rippin}}, \ and\
  \bibinfo {author} {\bibfnamefont {P.~L.}\ \bibnamefont {Knight}},\ }\href
  {\doibase 10.1103/PhysRevLett.78.2275} {\bibfield  {journal} {\bibinfo
  {journal} {Phys. Rev. Lett.}\ }\textbf {\bibinfo {volume} {78}},\ \bibinfo
  {pages} {2275} (\bibinfo {year} {1997})}\BibitemShut {NoStop}%
\bibitem [{\citenamefont {Horodecki}\ \emph {et~al.}(2009)\citenamefont
  {Horodecki}, \citenamefont {Horodecki}, \citenamefont {Horodecki},\ and\
  \citenamefont {Horodecki}}]{RevModPhys.81.865}%
  \BibitemOpen
  \bibfield  {author} {\bibinfo {author} {\bibfnamefont {R.}~\bibnamefont
  {Horodecki}}, \bibinfo {author} {\bibfnamefont {P.}~\bibnamefont
  {Horodecki}}, \bibinfo {author} {\bibfnamefont {M.}~\bibnamefont
  {Horodecki}}, \ and\ \bibinfo {author} {\bibfnamefont {K.}~\bibnamefont
  {Horodecki}},\ }\href {\doibase 10.1103/RevModPhys.81.865} {\bibfield
  {journal} {\bibinfo  {journal} {Rev. Mod. Phys.}\ }\textbf {\bibinfo {volume}
  {81}},\ \bibinfo {pages} {865} (\bibinfo {year} {2009})}\BibitemShut
  {NoStop}%
\bibitem [{\citenamefont {Veitch}\ \emph {et~al.}(2012)\citenamefont {Veitch},
  \citenamefont {Ferrie}, \citenamefont {Gross},\ and\ \citenamefont
  {Emerson}}]{veitch2012negative}%
  \BibitemOpen
  \bibfield  {author} {\bibinfo {author} {\bibfnamefont {V.}~\bibnamefont
  {Veitch}}, \bibinfo {author} {\bibfnamefont {C.}~\bibnamefont {Ferrie}},
  \bibinfo {author} {\bibfnamefont {D.}~\bibnamefont {Gross}}, \ and\ \bibinfo
  {author} {\bibfnamefont {J.}~\bibnamefont {Emerson}},\ }\href@noop {}
  {\bibfield  {journal} {\bibinfo  {journal} {New Journal of Physics}\ }\textbf
  {\bibinfo {volume} {14}},\ \bibinfo {pages} {113011} (\bibinfo {year}
  {2012})}\BibitemShut {NoStop}%
\bibitem [{\citenamefont {Veitch}\ \emph {et~al.}(2014)\citenamefont {Veitch},
  \citenamefont {Mousavian}, \citenamefont {Gottesman},\ and\ \citenamefont
  {Emerson}}]{veitch2014resource}%
  \BibitemOpen
  \bibfield  {author} {\bibinfo {author} {\bibfnamefont {V.}~\bibnamefont
  {Veitch}}, \bibinfo {author} {\bibfnamefont {S.~H.}\ \bibnamefont
  {Mousavian}}, \bibinfo {author} {\bibfnamefont {D.}~\bibnamefont
  {Gottesman}}, \ and\ \bibinfo {author} {\bibfnamefont {J.}~\bibnamefont
  {Emerson}},\ }\href@noop {} {\bibfield  {journal} {\bibinfo  {journal} {New
  Journal of Physics}\ }\textbf {\bibinfo {volume} {16}},\ \bibinfo {pages}
  {013009} (\bibinfo {year} {2014})}\BibitemShut {NoStop}%
\bibitem [{\citenamefont {Howard}\ and\ \citenamefont
  {Campbell}(2017)}]{PhysRevLett.118.090501}%
  \BibitemOpen
  \bibfield  {author} {\bibinfo {author} {\bibfnamefont {M.}~\bibnamefont
  {Howard}}\ and\ \bibinfo {author} {\bibfnamefont {E.}~\bibnamefont
  {Campbell}},\ }\href {\doibase 10.1103/PhysRevLett.118.090501} {\bibfield
  {journal} {\bibinfo  {journal} {Phys. Rev. Lett.}\ }\textbf {\bibinfo
  {volume} {118}},\ \bibinfo {pages} {090501} (\bibinfo {year}
  {2017})}\BibitemShut {NoStop}%
\bibitem [{\citenamefont {{Hsieh}}\ \emph {et~al.}(2017)\citenamefont
  {{Hsieh}}, \citenamefont {{Chen}},\ and\ \citenamefont
  {{Li}}}]{quantifying17sr}%
  \BibitemOpen
  \bibfield  {author} {\bibinfo {author} {\bibfnamefont {J.-H.}\ \bibnamefont
  {{Hsieh}}}, \bibinfo {author} {\bibfnamefont {S.-H.}\ \bibnamefont {{Chen}}},
  \ and\ \bibinfo {author} {\bibfnamefont {C.-M.}\ \bibnamefont {{Li}}},\
  }\href {\doibase 10.1038/s41598-017-13604-9} {\bibfield  {journal} {\bibinfo
  {journal} {Scientific Reports}\ }\textbf {\bibinfo {volume} {7}},\ \bibinfo
  {eid} {13588} (\bibinfo {year} {2017})},\ \Eprint
  {http://arxiv.org/abs/1710.07068} {arXiv:1710.07068 [quant-ph]} \BibitemShut
  {NoStop}%
\bibitem [{\citenamefont {D{\'{i}}az}\ \emph {et~al.}(2018)\citenamefont
  {D{\'{i}}az}, \citenamefont {Fang}, \citenamefont {Wang}, \citenamefont
  {Rosati}, \citenamefont {Skotiniotis}, \citenamefont {Calsamiglia},\ and\
  \citenamefont {Winter}}]{Diaz2018usingreusing}%
  \BibitemOpen
  \bibfield  {author} {\bibinfo {author} {\bibfnamefont {M.~G.}\ \bibnamefont
  {D{\'{i}}az}}, \bibinfo {author} {\bibfnamefont {K.}~\bibnamefont {Fang}},
  \bibinfo {author} {\bibfnamefont {X.}~\bibnamefont {Wang}}, \bibinfo {author}
  {\bibfnamefont {M.}~\bibnamefont {Rosati}}, \bibinfo {author} {\bibfnamefont
  {M.}~\bibnamefont {Skotiniotis}}, \bibinfo {author} {\bibfnamefont
  {J.}~\bibnamefont {Calsamiglia}}, \ and\ \bibinfo {author} {\bibfnamefont
  {A.}~\bibnamefont {Winter}},\ }\href {\doibase 10.22331/q-2018-10-19-100}
  {\bibfield  {journal} {\bibinfo  {journal} {{Quantum}}\ }\textbf {\bibinfo
  {volume} {2}},\ \bibinfo {pages} {100} (\bibinfo {year} {2018})}\BibitemShut
  {NoStop}%
\bibitem [{\citenamefont {Saxena}\ \emph {et~al.}(2020)\citenamefont {Saxena},
  \citenamefont {Chitambar},\ and\ \citenamefont {Gour}}]{dynamicalCoherence}%
  \BibitemOpen
  \bibfield  {author} {\bibinfo {author} {\bibfnamefont {G.}~\bibnamefont
  {Saxena}}, \bibinfo {author} {\bibfnamefont {E.}~\bibnamefont {Chitambar}}, \
  and\ \bibinfo {author} {\bibfnamefont {G.}~\bibnamefont {Gour}},\ }\href
  {\doibase 10.1103/PhysRevResearch.2.023298} {\bibfield  {journal} {\bibinfo
  {journal} {Phys. Rev. Research}\ }\textbf {\bibinfo {volume} {2}},\ \bibinfo
  {pages} {023298} (\bibinfo {year} {2020})}\BibitemShut {NoStop}%
\bibitem [{\citenamefont {Theurer}\ \emph {et~al.}(2019)\citenamefont
  {Theurer}, \citenamefont {Egloff}, \citenamefont {Zhang},\ and\ \citenamefont
  {Plenio}}]{theurer2018quantifying}%
  \BibitemOpen
  \bibfield  {author} {\bibinfo {author} {\bibfnamefont {T.}~\bibnamefont
  {Theurer}}, \bibinfo {author} {\bibfnamefont {D.}~\bibnamefont {Egloff}},
  \bibinfo {author} {\bibfnamefont {L.}~\bibnamefont {Zhang}}, \ and\ \bibinfo
  {author} {\bibfnamefont {M.~B.}\ \bibnamefont {Plenio}},\ }\href {\doibase
  10.1103/PhysRevLett.122.190405} {\bibfield  {journal} {\bibinfo  {journal}
  {Phys. Rev. Lett.}\ }\textbf {\bibinfo {volume} {122}},\ \bibinfo {pages}
  {190405} (\bibinfo {year} {2019})}\BibitemShut {NoStop}%
\bibitem [{\citenamefont {Theurer}\ \emph {et~al.}(2020)\citenamefont
  {Theurer}, \citenamefont {Satyajit},\ and\ \citenamefont
  {Plenio}}]{PhysRevLett.125.130401}%
  \BibitemOpen
  \bibfield  {author} {\bibinfo {author} {\bibfnamefont {T.}~\bibnamefont
  {Theurer}}, \bibinfo {author} {\bibfnamefont {S.}~\bibnamefont {Satyajit}}, \
  and\ \bibinfo {author} {\bibfnamefont {M.~B.}\ \bibnamefont {Plenio}},\
  }\href {\doibase 10.1103/PhysRevLett.125.130401} {\bibfield  {journal}
  {\bibinfo  {journal} {Phys. Rev. Lett.}\ }\textbf {\bibinfo {volume} {125}},\
  \bibinfo {pages} {130401} (\bibinfo {year} {2020})}\BibitemShut {NoStop}%
\bibitem [{\citenamefont {Rosset}\ \emph {et~al.}(2018)\citenamefont {Rosset},
  \citenamefont {Buscemi},\ and\ \citenamefont {Liang}}]{Rosset2018resource}%
  \BibitemOpen
  \bibfield  {author} {\bibinfo {author} {\bibfnamefont {D.}~\bibnamefont
  {Rosset}}, \bibinfo {author} {\bibfnamefont {F.}~\bibnamefont {Buscemi}}, \
  and\ \bibinfo {author} {\bibfnamefont {Y.-C.}\ \bibnamefont {Liang}},\ }\href
  {\doibase 10.1103/PhysRevX.8.021033} {\bibfield  {journal} {\bibinfo
  {journal} {Phys. Rev. X}\ }\textbf {\bibinfo {volume} {8}},\ \bibinfo {pages}
  {021033} (\bibinfo {year} {2018})}\BibitemShut {NoStop}%
\bibitem [{\citenamefont {Yuan}\ \emph {et~al.}(2019)\citenamefont {Yuan},
  \citenamefont {Liu}, \citenamefont {Zhao}, \citenamefont {Regula},
  \citenamefont {Thompson},\ and\ \citenamefont {Gu}}]{yuan2019robustness}%
  \BibitemOpen
  \bibfield  {author} {\bibinfo {author} {\bibfnamefont {X.}~\bibnamefont
  {Yuan}}, \bibinfo {author} {\bibfnamefont {Y.}~\bibnamefont {Liu}}, \bibinfo
  {author} {\bibfnamefont {Q.}~\bibnamefont {Zhao}}, \bibinfo {author}
  {\bibfnamefont {B.}~\bibnamefont {Regula}}, \bibinfo {author} {\bibfnamefont
  {J.}~\bibnamefont {Thompson}}, \ and\ \bibinfo {author} {\bibfnamefont
  {M.}~\bibnamefont {Gu}},\ }\href {https://arxiv.org/abs/1907.02521} {\enquote
  {\bibinfo {title} {Robustness of quantum memories: An operational
  resource-theoretic approach},}\ } (\bibinfo {year} {2019}),\ \Eprint
  {http://arxiv.org/abs/1907.02521} {arXiv:1907.02521 [quant-ph]} \BibitemShut
  {NoStop}%
\bibitem [{\citenamefont {Takagi}\ \emph {et~al.}(2020)\citenamefont {Takagi},
  \citenamefont {Wang},\ and\ \citenamefont {Hayashi}}]{takagi_2020}%
  \BibitemOpen
  \bibfield  {author} {\bibinfo {author} {\bibfnamefont {R.}~\bibnamefont
  {Takagi}}, \bibinfo {author} {\bibfnamefont {K.}~\bibnamefont {Wang}}, \ and\
  \bibinfo {author} {\bibfnamefont {M.}~\bibnamefont {Hayashi}},\ }\href
  {\doibase 10.1103/PhysRevLett.124.120502} {\bibfield  {journal} {\bibinfo
  {journal} {Phys. Rev. Lett.}\ }\textbf {\bibinfo {volume} {124}},\ \bibinfo
  {pages} {120502} (\bibinfo {year} {2020})}\BibitemShut {NoStop}%
\bibitem [{\citenamefont {{Kuo}}\ \emph {et~al.}(2018)\citenamefont {{Kuo}},
  \citenamefont {{Chen}}, \citenamefont {{Lee}}, \citenamefont {{Chen}},
  \citenamefont {{Lu}},\ and\ \citenamefont {{Li}}}]{2018arXiv181110307K}%
  \BibitemOpen
  \bibfield  {author} {\bibinfo {author} {\bibfnamefont {C.-C.}\ \bibnamefont
  {{Kuo}}}, \bibinfo {author} {\bibfnamefont {S.-H.}\ \bibnamefont {{Chen}}},
  \bibinfo {author} {\bibfnamefont {W.-T.}\ \bibnamefont {{Lee}}}, \bibinfo
  {author} {\bibfnamefont {H.-M.}\ \bibnamefont {{Chen}}}, \bibinfo {author}
  {\bibfnamefont {H.}~\bibnamefont {{Lu}}}, \ and\ \bibinfo {author}
  {\bibfnamefont {C.-M.}\ \bibnamefont {{Li}}},\ }\href@noop {} {\bibfield
  {journal} {\bibinfo  {journal} {``Quantum Process Capability'', arXiv
  preprint arXiv:1811.10307}\ } (\bibinfo {year} {2018})}\BibitemShut {NoStop}%
\bibitem [{\citenamefont {Zhuang}\ \emph {et~al.}(2018)\citenamefont {Zhuang},
  \citenamefont {Shor},\ and\ \citenamefont {Shapiro}}]{zhuang2018resource}%
  \BibitemOpen
  \bibfield  {author} {\bibinfo {author} {\bibfnamefont {Q.}~\bibnamefont
  {Zhuang}}, \bibinfo {author} {\bibfnamefont {P.~W.}\ \bibnamefont {Shor}}, \
  and\ \bibinfo {author} {\bibfnamefont {J.~H.}\ \bibnamefont {Shapiro}},\
  }\href {\doibase 10.1103/PhysRevA.97.052317} {\bibfield  {journal} {\bibinfo
  {journal} {Phys. Rev. A}\ }\textbf {\bibinfo {volume} {97}},\ \bibinfo
  {pages} {052317} (\bibinfo {year} {2018})}\BibitemShut {NoStop}%
\bibitem [{\citenamefont {Seddon}\ and\ \citenamefont
  {Campbell}(2019)}]{seddon2019quantifying}%
  \BibitemOpen
  \bibfield  {author} {\bibinfo {author} {\bibfnamefont {J.~R.}\ \bibnamefont
  {Seddon}}\ and\ \bibinfo {author} {\bibfnamefont {E.~T.}\ \bibnamefont
  {Campbell}},\ }\href {\doibase 10.1098/rspa.2019.0251} {\bibfield  {journal}
  {\bibinfo  {journal} {Proceedings of the Royal Society A: Mathematical,
  Physical and Engineering Sciences}\ }\textbf {\bibinfo {volume} {475}},\
  \bibinfo {pages} {20190251} (\bibinfo {year} {2019})}\BibitemShut {NoStop}%
\bibitem [{\citenamefont {Wang}\ \emph {et~al.}(2019)\citenamefont {Wang},
  \citenamefont {Wilde},\ and\ \citenamefont {Su}}]{wang2019quantifying}%
  \BibitemOpen
  \bibfield  {author} {\bibinfo {author} {\bibfnamefont {X.}~\bibnamefont
  {Wang}}, \bibinfo {author} {\bibfnamefont {M.~M.}\ \bibnamefont {Wilde}}, \
  and\ \bibinfo {author} {\bibfnamefont {Y.}~\bibnamefont {Su}},\ }\href
  {\doibase 10.1088/1367-2630/ab451d} {\bibfield  {journal} {\bibinfo
  {journal} {New J. Phys.}\ }\textbf {\bibinfo {volume} {21}},\ \bibinfo
  {pages} {103002} (\bibinfo {year} {2019})}\BibitemShut {NoStop}%
\bibitem [{\citenamefont {Takagi}\ and\ \citenamefont
  {Regula}(2019)}]{takagi2019general}%
  \BibitemOpen
  \bibfield  {author} {\bibinfo {author} {\bibfnamefont {R.}~\bibnamefont
  {Takagi}}\ and\ \bibinfo {author} {\bibfnamefont {B.}~\bibnamefont
  {Regula}},\ }\href {\doibase 10.1103/PhysRevX.9.031053} {\bibfield  {journal}
  {\bibinfo  {journal} {Phys. Rev. X}\ }\textbf {\bibinfo {volume} {9}},\
  \bibinfo {pages} {031053} (\bibinfo {year} {2019})}\BibitemShut {NoStop}%
\bibitem [{\citenamefont {Skrzypczyk}\ and\ \citenamefont
  {Linden}(2019)}]{PhysRevLett.122.140403}%
  \BibitemOpen
  \bibfield  {author} {\bibinfo {author} {\bibfnamefont {P.}~\bibnamefont
  {Skrzypczyk}}\ and\ \bibinfo {author} {\bibfnamefont {N.}~\bibnamefont
  {Linden}},\ }\href {\doibase 10.1103/PhysRevLett.122.140403} {\bibfield
  {journal} {\bibinfo  {journal} {Phys. Rev. Lett.}\ }\textbf {\bibinfo
  {volume} {122}},\ \bibinfo {pages} {140403} (\bibinfo {year}
  {2019})}\BibitemShut {NoStop}%
\bibitem [{\citenamefont {Liu}\ and\ \citenamefont
  {Yuan}(2020)}]{liu2019operational}%
  \BibitemOpen
  \bibfield  {author} {\bibinfo {author} {\bibfnamefont {Y.}~\bibnamefont
  {Liu}}\ and\ \bibinfo {author} {\bibfnamefont {X.}~\bibnamefont {Yuan}},\
  }\href {\doibase 10.1103/PhysRevResearch.2.012035} {\bibfield  {journal}
  {\bibinfo  {journal} {Phys. Rev. Research}\ }\textbf {\bibinfo {volume}
  {2}},\ \bibinfo {pages} {012035} (\bibinfo {year} {2020})}\BibitemShut
  {NoStop}%
\bibitem [{\citenamefont {Liu}\ and\ \citenamefont
  {Winter}(2019)}]{liu2019resource}%
  \BibitemOpen
  \bibfield  {author} {\bibinfo {author} {\bibfnamefont {Z.-W.}\ \bibnamefont
  {Liu}}\ and\ \bibinfo {author} {\bibfnamefont {A.}~\bibnamefont {Winter}},\
  }\href@noop {} {\bibfield  {journal} {\bibinfo  {journal} {``Resource
  theories of quantum channels and the universal role of resource erasure'',
  arXiv preprint arXiv:1904.04201}\ } (\bibinfo {year} {2019})}\BibitemShut
  {NoStop}%
\bibitem [{\citenamefont {{Fang}}\ and\ \citenamefont
  {{Liu}}(2020)}]{2020arXiv201011822F}%
  \BibitemOpen
  \bibfield  {author} {\bibinfo {author} {\bibfnamefont {K.}~\bibnamefont
  {{Fang}}}\ and\ \bibinfo {author} {\bibfnamefont {Z.-W.}\ \bibnamefont
  {{Liu}}},\ }\href@noop {} {\bibfield  {journal} {\bibinfo  {journal} {arXiv
  e-prints}\ ,\ \bibinfo {eid} {arXiv:2010.11822}} (\bibinfo {year} {2020})},\
  \Eprint {http://arxiv.org/abs/2010.11822} {arXiv:2010.11822 [quant-ph]}
  \BibitemShut {NoStop}%
\bibitem [{\citenamefont {{Regula}}\ and\ \citenamefont
  {{Takagi}}(2020)}]{2020arXiv201011942R}%
  \BibitemOpen
  \bibfield  {author} {\bibinfo {author} {\bibfnamefont {B.}~\bibnamefont
  {{Regula}}}\ and\ \bibinfo {author} {\bibfnamefont {R.}~\bibnamefont
  {{Takagi}}},\ }\href@noop {} {\bibfield  {journal} {\bibinfo  {journal}
  {arXiv e-prints}\ ,\ \bibinfo {eid} {arXiv:2010.11942}} (\bibinfo {year}
  {2020})},\ \Eprint {http://arxiv.org/abs/2010.11942} {arXiv:2010.11942
  [quant-ph]} \BibitemShut {NoStop}%
\bibitem [{\citenamefont {Wang}\ and\ \citenamefont
  {Wilde}(2019)}]{PhysRevResearch.1.033169}%
  \BibitemOpen
  \bibfield  {author} {\bibinfo {author} {\bibfnamefont {X.}~\bibnamefont
  {Wang}}\ and\ \bibinfo {author} {\bibfnamefont {M.~M.}\ \bibnamefont
  {Wilde}},\ }\href {\doibase 10.1103/PhysRevResearch.1.033169} {\bibfield
  {journal} {\bibinfo  {journal} {Phys. Rev. Research}\ }\textbf {\bibinfo
  {volume} {1}},\ \bibinfo {pages} {033169} (\bibinfo {year}
  {2019})}\BibitemShut {NoStop}%
\bibitem [{\citenamefont {Takagi}(2020)}]{takagi2020optimal}%
  \BibitemOpen
  \bibfield  {author} {\bibinfo {author} {\bibfnamefont {R.}~\bibnamefont
  {Takagi}},\ }\href@noop {} {\  (\bibinfo {year} {2020})},\ \Eprint
  {http://arxiv.org/abs/2006.12509} {arXiv:2006.12509 [quant-ph]} \BibitemShut
  {NoStop}%
\bibitem [{\citenamefont {Bauml}\ \emph {et~al.}(2019)\citenamefont {Bauml},
  \citenamefont {Das}, \citenamefont {Wang},\ and\ \citenamefont
  {Wilde}}]{bauml2019resource}%
  \BibitemOpen
  \bibfield  {author} {\bibinfo {author} {\bibfnamefont {S.}~\bibnamefont
  {Bauml}}, \bibinfo {author} {\bibfnamefont {S.}~\bibnamefont {Das}}, \bibinfo
  {author} {\bibfnamefont {X.}~\bibnamefont {Wang}}, \ and\ \bibinfo {author}
  {\bibfnamefont {M.~M.}\ \bibnamefont {Wilde}},\ }\href@noop {} {\enquote
  {\bibinfo {title} {Resource theory of entanglement for bipartite quantum
  channels},}\ } (\bibinfo {year} {2019}),\ \Eprint
  {http://arxiv.org/abs/1907.04181} {arXiv:1907.04181 [quant-ph]} \BibitemShut
  {NoStop}%
\bibitem [{\citenamefont {Gour}\ and\ \citenamefont
  {Scandolo}(2020)}]{dynamicalEntanglement}%
  \BibitemOpen
  \bibfield  {author} {\bibinfo {author} {\bibfnamefont {G.}~\bibnamefont
  {Gour}}\ and\ \bibinfo {author} {\bibfnamefont {C.~M.}\ \bibnamefont
  {Scandolo}},\ }\href {\doibase 10.1103/PhysRevLett.125.180505} {\bibfield
  {journal} {\bibinfo  {journal} {Phys. Rev. Lett.}\ }\textbf {\bibinfo
  {volume} {125}},\ \bibinfo {pages} {180505} (\bibinfo {year}
  {2020})}\BibitemShut {NoStop}%
\bibitem [{\citenamefont {Wang}\ and\ \citenamefont
  {Wilde}(2020)}]{PhysRevLett.125.040502}%
  \BibitemOpen
  \bibfield  {author} {\bibinfo {author} {\bibfnamefont {X.}~\bibnamefont
  {Wang}}\ and\ \bibinfo {author} {\bibfnamefont {M.~M.}\ \bibnamefont
  {Wilde}},\ }\href {\doibase 10.1103/PhysRevLett.125.040502} {\bibfield
  {journal} {\bibinfo  {journal} {Phys. Rev. Lett.}\ }\textbf {\bibinfo
  {volume} {125}},\ \bibinfo {pages} {040502} (\bibinfo {year}
  {2020})}\BibitemShut {NoStop}%
\bibitem [{\citenamefont {Chiribella}\ \emph {et~al.}(2008)\citenamefont
  {Chiribella}, \citenamefont {Ariano},\ and\ \citenamefont
  {Perinotti}}]{chiribella2008transform}%
  \BibitemOpen
  \bibfield  {author} {\bibinfo {author} {\bibfnamefont {G.}~\bibnamefont
  {Chiribella}}, \bibinfo {author} {\bibfnamefont {G.~M.~D.}\ \bibnamefont
  {Ariano}}, \ and\ \bibinfo {author} {\bibfnamefont {P.}~\bibnamefont
  {Perinotti}},\ }\href {\doibase 10.1209/0295-5075/83/30004} {\bibfield
  {journal} {\bibinfo  {journal} {{EPL} (Europhysics Letters)}\ }\textbf
  {\bibinfo {volume} {83}},\ \bibinfo {pages} {30004} (\bibinfo {year}
  {2008})}\BibitemShut {NoStop}%
\bibitem [{\citenamefont {Gour}(2019)}]{gour2019comparison}%
  \BibitemOpen
  \bibfield  {author} {\bibinfo {author} {\bibfnamefont {G.}~\bibnamefont
  {Gour}},\ }\href {https://ieeexplore.ieee.org/abstract/document/8678741}
  {\bibfield  {journal} {\bibinfo  {journal} {IEEE Transactions on Information
  Theory}\ }\textbf {\bibinfo {volume} {65}},\ \bibinfo {pages} {5880}
  (\bibinfo {year} {2019})}\BibitemShut {NoStop}%
\bibitem [{\citenamefont {Mani}\ and\ \citenamefont
  {Karimipour}(2015)}]{Mani2015cohering}%
  \BibitemOpen
  \bibfield  {author} {\bibinfo {author} {\bibfnamefont {A.}~\bibnamefont
  {Mani}}\ and\ \bibinfo {author} {\bibfnamefont {V.}~\bibnamefont
  {Karimipour}},\ }\href {\doibase 10.1103/PhysRevA.92.032331} {\bibfield
  {journal} {\bibinfo  {journal} {Phys. Rev. A}\ }\textbf {\bibinfo {volume}
  {92}},\ \bibinfo {pages} {032331} (\bibinfo {year} {2015})}\BibitemShut
  {NoStop}%
\bibitem [{\citenamefont {Zanardi}\ \emph
  {et~al.}(2017{\natexlab{a}})\citenamefont {Zanardi}, \citenamefont
  {Styliaris},\ and\ \citenamefont {Campos~Venuti}}]{Zanardi2017measures}%
  \BibitemOpen
  \bibfield  {author} {\bibinfo {author} {\bibfnamefont {P.}~\bibnamefont
  {Zanardi}}, \bibinfo {author} {\bibfnamefont {G.}~\bibnamefont {Styliaris}},
  \ and\ \bibinfo {author} {\bibfnamefont {L.}~\bibnamefont {Campos~Venuti}},\
  }\href {\doibase 10.1103/PhysRevA.95.052307} {\bibfield  {journal} {\bibinfo
  {journal} {Phys. Rev. A}\ }\textbf {\bibinfo {volume} {95}},\ \bibinfo
  {pages} {052307} (\bibinfo {year} {2017}{\natexlab{a}})}\BibitemShut
  {NoStop}%
\bibitem [{\citenamefont {Zanardi}\ \emph
  {et~al.}(2017{\natexlab{b}})\citenamefont {Zanardi}, \citenamefont
  {Styliaris},\ and\ \citenamefont {Campos~Venuti}}]{Zanardi2017coherence}%
  \BibitemOpen
  \bibfield  {author} {\bibinfo {author} {\bibfnamefont {P.}~\bibnamefont
  {Zanardi}}, \bibinfo {author} {\bibfnamefont {G.}~\bibnamefont {Styliaris}},
  \ and\ \bibinfo {author} {\bibfnamefont {L.}~\bibnamefont {Campos~Venuti}},\
  }\href {\doibase 10.1103/PhysRevA.95.052306} {\bibfield  {journal} {\bibinfo
  {journal} {Phys. Rev. A}\ }\textbf {\bibinfo {volume} {95}},\ \bibinfo
  {pages} {052306} (\bibinfo {year} {2017}{\natexlab{b}})}\BibitemShut
  {NoStop}%
\bibitem [{\citenamefont {Bu}\ \emph {et~al.}(2017)\citenamefont {Bu},
  \citenamefont {Kumar}, \citenamefont {Zhang},\ and\ \citenamefont
  {Wu}}]{BU20171670}%
  \BibitemOpen
  \bibfield  {author} {\bibinfo {author} {\bibfnamefont {K.}~\bibnamefont
  {Bu}}, \bibinfo {author} {\bibfnamefont {A.}~\bibnamefont {Kumar}}, \bibinfo
  {author} {\bibfnamefont {L.}~\bibnamefont {Zhang}}, \ and\ \bibinfo {author}
  {\bibfnamefont {J.}~\bibnamefont {Wu}},\ }\href {\doibase
  https://doi.org/10.1016/j.physleta.2017.03.022} {\bibfield  {journal}
  {\bibinfo  {journal} {Physics Letters A}\ }\textbf {\bibinfo {volume}
  {381}},\ \bibinfo {pages} {1670 } (\bibinfo {year} {2017})}\BibitemShut
  {NoStop}%
\bibitem [{\citenamefont {Ben~Dana}\ \emph {et~al.}(2017)\citenamefont
  {Ben~Dana}, \citenamefont {Garc\'{\i}a~D\'{\i}az}, \citenamefont {Mejatty},\
  and\ \citenamefont {Winter}}]{Dana2018resource}%
  \BibitemOpen
  \bibfield  {author} {\bibinfo {author} {\bibfnamefont {K.}~\bibnamefont
  {Ben~Dana}}, \bibinfo {author} {\bibfnamefont {M.}~\bibnamefont
  {Garc\'{\i}a~D\'{\i}az}}, \bibinfo {author} {\bibfnamefont {M.}~\bibnamefont
  {Mejatty}}, \ and\ \bibinfo {author} {\bibfnamefont {A.}~\bibnamefont
  {Winter}},\ }\href {\doibase 10.1103/PhysRevA.95.062327} {\bibfield
  {journal} {\bibinfo  {journal} {Phys. Rev. A}\ }\textbf {\bibinfo {volume}
  {95}},\ \bibinfo {pages} {062327} (\bibinfo {year} {2017})}\BibitemShut
  {NoStop}%
\bibitem [{\citenamefont {Regula}\ \emph {et~al.}(2020)\citenamefont {Regula},
  \citenamefont {Bu}, \citenamefont {Takagi},\ and\ \citenamefont
  {Liu}}]{PhysRevA.101.062315}%
  \BibitemOpen
  \bibfield  {author} {\bibinfo {author} {\bibfnamefont {B.}~\bibnamefont
  {Regula}}, \bibinfo {author} {\bibfnamefont {K.}~\bibnamefont {Bu}}, \bibinfo
  {author} {\bibfnamefont {R.}~\bibnamefont {Takagi}}, \ and\ \bibinfo {author}
  {\bibfnamefont {Z.-W.}\ \bibnamefont {Liu}},\ }\href {\doibase
  10.1103/PhysRevA.101.062315} {\bibfield  {journal} {\bibinfo  {journal}
  {Phys. Rev. A}\ }\textbf {\bibinfo {volume} {101}},\ \bibinfo {pages}
  {062315} (\bibinfo {year} {2020})}\BibitemShut {NoStop}%
\bibitem [{\citenamefont {Fang}\ and\ \citenamefont
  {Liu}(2020)}]{PhysRevLett.125.060405}%
  \BibitemOpen
  \bibfield  {author} {\bibinfo {author} {\bibfnamefont {K.}~\bibnamefont
  {Fang}}\ and\ \bibinfo {author} {\bibfnamefont {Z.-W.}\ \bibnamefont {Liu}},\
  }\href {\doibase 10.1103/PhysRevLett.125.060405} {\bibfield  {journal}
  {\bibinfo  {journal} {Phys. Rev. Lett.}\ }\textbf {\bibinfo {volume} {125}},\
  \bibinfo {pages} {060405} (\bibinfo {year} {2020})}\BibitemShut {NoStop}%
\bibitem [{\citenamefont {Liu}\ \emph {et~al.}(2019)\citenamefont {Liu},
  \citenamefont {Bu},\ and\ \citenamefont {Takagi}}]{2019arXiv190405840L}%
  \BibitemOpen
  \bibfield  {author} {\bibinfo {author} {\bibfnamefont {Z.-W.}\ \bibnamefont
  {Liu}}, \bibinfo {author} {\bibfnamefont {K.}~\bibnamefont {Bu}}, \ and\
  \bibinfo {author} {\bibfnamefont {R.}~\bibnamefont {Takagi}},\ }\href
  {\doibase 10.1103/PhysRevLett.123.020401} {\bibfield  {journal} {\bibinfo
  {journal} {Phys. Rev. Lett.}\ }\textbf {\bibinfo {volume} {123}},\ \bibinfo
  {pages} {020401} (\bibinfo {year} {2019})}\BibitemShut {NoStop}%
\bibitem [{\citenamefont {Gour}\ and\ \citenamefont
  {Winter}(2019)}]{PhysRevLett.123.150401}%
  \BibitemOpen
  \bibfield  {author} {\bibinfo {author} {\bibfnamefont {G.}~\bibnamefont
  {Gour}}\ and\ \bibinfo {author} {\bibfnamefont {A.}~\bibnamefont {Winter}},\
  }\href {\doibase 10.1103/PhysRevLett.123.150401} {\bibfield  {journal}
  {\bibinfo  {journal} {Phys. Rev. Lett.}\ }\textbf {\bibinfo {volume} {123}},\
  \bibinfo {pages} {150401} (\bibinfo {year} {2019})}\BibitemShut {NoStop}%
\bibitem [{\citenamefont {Aberg}(2006)}]{aberg2006quantifying}%
  \BibitemOpen
  \bibfield  {author} {\bibinfo {author} {\bibfnamefont {J.}~\bibnamefont
  {Aberg}},\ }\href@noop {} {\bibfield  {journal} {\bibinfo  {journal} {arXiv
  preprint quant-ph/0612146}\ } (\bibinfo {year} {2006})}\BibitemShut {NoStop}%
\bibitem [{\citenamefont {Baumgratz}\ \emph {et~al.}(2014)\citenamefont
  {Baumgratz}, \citenamefont {Cramer},\ and\ \citenamefont
  {Plenio}}]{Baumgratz14}%
  \BibitemOpen
  \bibfield  {author} {\bibinfo {author} {\bibfnamefont {T.}~\bibnamefont
  {Baumgratz}}, \bibinfo {author} {\bibfnamefont {M.}~\bibnamefont {Cramer}}, \
  and\ \bibinfo {author} {\bibfnamefont {M.~B.}\ \bibnamefont {Plenio}},\
  }\href {\doibase 10.1103/PhysRevLett.113.140401} {\bibfield  {journal}
  {\bibinfo  {journal} {Phys. Rev. Lett.}\ }\textbf {\bibinfo {volume} {113}},\
  \bibinfo {pages} {140401} (\bibinfo {year} {2014})}\BibitemShut {NoStop}%
\bibitem [{\citenamefont {Streltsov}\ \emph {et~al.}(2017)\citenamefont
  {Streltsov}, \citenamefont {Adesso},\ and\ \citenamefont
  {Plenio}}]{RevModPhys.89.041003}%
  \BibitemOpen
  \bibfield  {author} {\bibinfo {author} {\bibfnamefont {A.}~\bibnamefont
  {Streltsov}}, \bibinfo {author} {\bibfnamefont {G.}~\bibnamefont {Adesso}}, \
  and\ \bibinfo {author} {\bibfnamefont {M.~B.}\ \bibnamefont {Plenio}},\
  }\href {\doibase 10.1103/RevModPhys.89.041003} {\bibfield  {journal}
  {\bibinfo  {journal} {Rev. Mod. Phys.}\ }\textbf {\bibinfo {volume} {89}},\
  \bibinfo {pages} {041003} (\bibinfo {year} {2017})}\BibitemShut {NoStop}%
\bibitem [{\citenamefont {Chitambar}\ and\ \citenamefont
  {Gour}(2016)}]{Chitambar16prl}%
  \BibitemOpen
  \bibfield  {author} {\bibinfo {author} {\bibfnamefont {E.}~\bibnamefont
  {Chitambar}}\ and\ \bibinfo {author} {\bibfnamefont {G.}~\bibnamefont
  {Gour}},\ }\href {\doibase 10.1103/PhysRevLett.117.030401} {\bibfield
  {journal} {\bibinfo  {journal} {Phys. Rev. Lett.}\ }\textbf {\bibinfo
  {volume} {117}},\ \bibinfo {pages} {030401} (\bibinfo {year}
  {2016})}\BibitemShut {NoStop}%
\bibitem [{\citenamefont {Marvian}\ and\ \citenamefont
  {Spekkens}(2016)}]{Marvian16}%
  \BibitemOpen
  \bibfield  {author} {\bibinfo {author} {\bibfnamefont {I.}~\bibnamefont
  {Marvian}}\ and\ \bibinfo {author} {\bibfnamefont {R.~W.}\ \bibnamefont
  {Spekkens}},\ }\href {\doibase 10.1103/PhysRevA.94.052324} {\bibfield
  {journal} {\bibinfo  {journal} {Phys. Rev. A}\ }\textbf {\bibinfo {volume}
  {94}},\ \bibinfo {pages} {052324} (\bibinfo {year} {2016})}\BibitemShut
  {NoStop}%
\bibitem [{\citenamefont {Winter}\ and\ \citenamefont {Yang}(2016)}]{Winter16}%
  \BibitemOpen
  \bibfield  {author} {\bibinfo {author} {\bibfnamefont {A.}~\bibnamefont
  {Winter}}\ and\ \bibinfo {author} {\bibfnamefont {D.}~\bibnamefont {Yang}},\
  }\href {\doibase 10.1103/PhysRevLett.116.120404} {\bibfield  {journal}
  {\bibinfo  {journal} {Phys. Rev. Lett.}\ }\textbf {\bibinfo {volume} {116}},\
  \bibinfo {pages} {120404} (\bibinfo {year} {2016})}\BibitemShut {NoStop}%
\bibitem [{\citenamefont {Zhao}\ \emph
  {et~al.}(2018{\natexlab{a}})\citenamefont {Zhao}, \citenamefont {Liu},
  \citenamefont {Yuan}, \citenamefont {Chitambar},\ and\ \citenamefont
  {Ma}}]{zhao2018oneshot}%
  \BibitemOpen
  \bibfield  {author} {\bibinfo {author} {\bibfnamefont {Q.}~\bibnamefont
  {Zhao}}, \bibinfo {author} {\bibfnamefont {Y.}~\bibnamefont {Liu}}, \bibinfo
  {author} {\bibfnamefont {X.}~\bibnamefont {Yuan}}, \bibinfo {author}
  {\bibfnamefont {E.}~\bibnamefont {Chitambar}}, \ and\ \bibinfo {author}
  {\bibfnamefont {X.}~\bibnamefont {Ma}},\ }\href {\doibase
  10.1103/PhysRevLett.120.070403} {\bibfield  {journal} {\bibinfo  {journal}
  {Phys. Rev. Lett.}\ }\textbf {\bibinfo {volume} {120}},\ \bibinfo {pages}
  {070403} (\bibinfo {year} {2018}{\natexlab{a}})}\BibitemShut {NoStop}%
\bibitem [{\citenamefont {Zhao}\ \emph
  {et~al.}(2018{\natexlab{b}})\citenamefont {Zhao}, \citenamefont {Liu},
  \citenamefont {Yuan}, \citenamefont {Chitambar},\ and\ \citenamefont
  {Winter}}]{zhao2018oneshotdistill}%
  \BibitemOpen
  \bibfield  {author} {\bibinfo {author} {\bibfnamefont {Q.}~\bibnamefont
  {Zhao}}, \bibinfo {author} {\bibfnamefont {Y.}~\bibnamefont {Liu}}, \bibinfo
  {author} {\bibfnamefont {X.}~\bibnamefont {Yuan}}, \bibinfo {author}
  {\bibfnamefont {E.}~\bibnamefont {Chitambar}}, \ and\ \bibinfo {author}
  {\bibfnamefont {A.}~\bibnamefont {Winter}},\ }\href@noop {} {\bibfield
  {journal} {\bibinfo  {journal} {arXiv preprint arXiv:1808.01885}\ } (\bibinfo
  {year} {2018}{\natexlab{b}})}\BibitemShut {NoStop}%
\bibitem [{\citenamefont {Regula}\ \emph {et~al.}(2018)\citenamefont {Regula},
  \citenamefont {Fang}, \citenamefont {Wang},\ and\ \citenamefont
  {Adesso}}]{regula2018one}%
  \BibitemOpen
  \bibfield  {author} {\bibinfo {author} {\bibfnamefont {B.}~\bibnamefont
  {Regula}}, \bibinfo {author} {\bibfnamefont {K.}~\bibnamefont {Fang}},
  \bibinfo {author} {\bibfnamefont {X.}~\bibnamefont {Wang}}, \ and\ \bibinfo
  {author} {\bibfnamefont {G.}~\bibnamefont {Adesso}},\ }\href {\doibase
  10.1103/PhysRevLett.121.010401} {\bibfield  {journal} {\bibinfo  {journal}
  {Phys. Rev. Lett.}\ }\textbf {\bibinfo {volume} {121}},\ \bibinfo {pages}
  {010401} (\bibinfo {year} {2018})}\BibitemShut {NoStop}%
\bibitem [{\citenamefont {Yuan}(2019)}]{yuan2019hypothesis}%
  \BibitemOpen
  \bibfield  {author} {\bibinfo {author} {\bibfnamefont {X.}~\bibnamefont
  {Yuan}},\ }\href@noop {} {\bibfield  {journal} {\bibinfo  {journal} {Physical
  Review A}\ }\textbf {\bibinfo {volume} {99}},\ \bibinfo {pages} {032317}
  (\bibinfo {year} {2019})}\BibitemShut {NoStop}%
\bibitem [{\citenamefont {Gour}\ and\ \citenamefont
  {Wilde}(2018)}]{gour2018entropy}%
  \BibitemOpen
  \bibfield  {author} {\bibinfo {author} {\bibfnamefont {G.}~\bibnamefont
  {Gour}}\ and\ \bibinfo {author} {\bibfnamefont {M.~M.}\ \bibnamefont
  {Wilde}},\ }\href {https://arxiv.org/abs/1808.06980} {\bibfield  {journal}
  {\bibinfo  {journal} {arXiv preprint arXiv:1808.06980}\ } (\bibinfo {year}
  {2018})}\BibitemShut {NoStop}%
\bibitem [{\citenamefont {Fang}\ \emph {et~al.}(2020)\citenamefont {Fang},
  \citenamefont {Wang}, \citenamefont {Tomamichel},\ and\ \citenamefont
  {Berta}}]{Fang_2020}%
  \BibitemOpen
  \bibfield  {author} {\bibinfo {author} {\bibfnamefont {K.}~\bibnamefont
  {Fang}}, \bibinfo {author} {\bibfnamefont {X.}~\bibnamefont {Wang}}, \bibinfo
  {author} {\bibfnamefont {M.}~\bibnamefont {Tomamichel}}, \ and\ \bibinfo
  {author} {\bibfnamefont {M.}~\bibnamefont {Berta}},\ }\href {\doibase
  10.1109/tit.2019.2943858} {\bibfield  {journal} {\bibinfo  {journal} {IEEE
  Transactions on Information Theory}\ }\textbf {\bibinfo {volume} {66}},\
  \bibinfo {pages} {2129–2140} (\bibinfo {year} {2020})}\BibitemShut
  {NoStop}%
\bibitem [{\citenamefont {Gour}\ \emph {et~al.}(2015)\citenamefont {Gour},
  \citenamefont {Müller}, \citenamefont {Narasimhachar}, \citenamefont
  {Spekkens},\ and\ \citenamefont {Yunger~Halpern}}]{Gour_2015}%
  \BibitemOpen
  \bibfield  {author} {\bibinfo {author} {\bibfnamefont {G.}~\bibnamefont
  {Gour}}, \bibinfo {author} {\bibfnamefont {M.~P.}\ \bibnamefont {Müller}},
  \bibinfo {author} {\bibfnamefont {V.}~\bibnamefont {Narasimhachar}}, \bibinfo
  {author} {\bibfnamefont {R.~W.}\ \bibnamefont {Spekkens}}, \ and\ \bibinfo
  {author} {\bibfnamefont {N.}~\bibnamefont {Yunger~Halpern}},\ }\href
  {\doibase 10.1016/j.physrep.2015.04.003} {\bibfield  {journal} {\bibinfo
  {journal} {Physics Reports}\ }\textbf {\bibinfo {volume} {583}},\ \bibinfo
  {pages} {1–58} (\bibinfo {year} {2015})}\BibitemShut {NoStop}%
\bibitem [{\citenamefont {Regula1}\ and\ \citenamefont
  {Takagi1}(2020)}]{dynamicalBartosz}%
  \BibitemOpen
  \bibfield  {author} {\bibinfo {author} {\bibfnamefont {B.}~\bibnamefont
  {Regula1}}\ and\ \bibinfo {author} {\bibfnamefont {R.}~\bibnamefont
  {Takagi1}},\ }\href@noop {} {\bibfield  {journal} {\bibinfo  {journal}
  {appearing on the arXiv concurrently}\ } (\bibinfo {year}
  {2020})}\BibitemShut {NoStop}%
\end{thebibliography}

%

\end{document}